\documentclass[conf]{new-aiaa}
\usepackage[utf8]{inputenc}
\usepackage[toc,acronym,nonumberlist]{glossaries}
\usepackage{graphicx}
\usepackage{amsmath}
\usepackage[version=4]{mhchem}
\usepackage{siunitx}
\usepackage{multicol}
\usepackage{booktabs}
 \usepackage{float}
\usepackage{textgreek}
\usepackage[ruled,vlined]{algorithm2e}
\usepackage{longtable,tabularx}
\usepackage{tabu}
\usepackage{tikz}
\usetikzlibrary{matrix,chains,positioning,decorations.pathreplacing,arrows}
\usepackage{subcaption}
\usetikzlibrary{decorations.pathreplacing}
\usetikzlibrary{fadings}
\newcommand\fithere{9.4cm}
\newcommand\rl{\acrshort{RL} }
\newcommand\sac{\acrshort{SAC} }

\newcommand\thetaparam{\text{\boldmath$\theta$}}

\title{Soft Actor-Critic Deep Reinforcement Learning for Fault-Tolerant Flight Control}

\author{Killian Dally\footnote{M.Sc., Faculty of Aerospace Engineering, Control and Simulation Division, Delft University of Technology}, Erik-Jan van Kampen\footnote{Assistant Professor, Faculty of Aerospace Engineering, Control and Simulation Division, Delft University of Technology}}
\affil{Delft University of Technology, P.O. Box 5058, 2600GB Delft, The Netherlands}

\newacronym{RL}{RL}{Reinforcement Learning}
\newacronym{DRL}{DRL}{Deep Reinforcement Learning}
\newacronym{TD}{TD}{Temporal Difference}

\newacronym{DP}{DP}{Dynamic Programming}

\newacronym{MC}{MC}{Monte-Carlo Method}
\newacronym{DASMAT}{DASMAT}{Delft University Aircraft Simulation Model and Analysis Tool}
\newacronym{EMA}{EMA}{Exponentially (Weighted) Moving Average}
\newacronym{VTOL}{VTOL}{Vertical Take-Off and Landing}
\newacronym{NN}{NN}{Neural Network}
\newacronym{DNN}{DNN}{Deep Neural Network}

\newacronym{ANDI}{ANDI}{Adaptive Nonlinear Dynamic Inversion}
\newacronym{INDI}{INDI}{Incremental Nonlinear Dynamic Inversion}
\newacronym{ABS}{ABS}{Adaptive Backstepping}

\newacronym{iADP}{iADP}{Incremental Approximate Dynamic Programming}

\newacronym{TD3}{TD3}{Twin-Delayed Deep Deterministic Policy Gradient}

\newacronym{DDPG}{DDPG}{Deep Deterministic Policy Gradient}

\newacronym{DPG}{DPG}{Deterministic Policy Gradient}

\newacronym{IFC}{IFC}{Initial Flight Conditions}

\newacronym{nMAE}{nMAE}{normalized Mean Absolute Error}

\newacronym{FDD}{FDD}{Fault Diagnosis and Detection}

\newacronym{UAV}{UAV}{Unmanned Air Vehicle}

\newacronym{ADP}{ADP}{Approximate Dynamic Programming}
\newacronym{MIMO}{MIMO}{Multi-Input Multi-Output}
\newacronym{SISO}{SISO}{Single-Input Single-Output}

\newacronym{HDP}{HDP}{Heuristic Dynamic Programming}
\newacronym{IHDP}{IHDP}{Incremental Heuristic Dynamic Programming}
\newacronym{IDHP}{IDHP}{Incremental Dual Heuristic Programming}
\newacronym{DHP}{DHP}{Dual Heuristic Programming}
\newacronym{GDHP}{GDHP}{Global Dual Heuristic Programming}
\newacronym{IGDHP}{IGDHP}{Incremental Global Dual Heuristic Programming}

\newacronym{ADHDP}{ADHDP}{Action-Dependent Heuristic Dynamic Programming}
\newacronym{ADDHP}{ADDHP}{Action-Dependent Dual Heuristic Programming}
\newacronym{ADGDHP}{ADGDHP}{Action-Dependent Global Dual Heuristic Programming}

\newacronym{i.i.d.}{i.i.d.}{Independent and Identically-Distributed}
\newacronym{AD}{AD}{Action-Dependent}

\newacronym{DoF}{DoF}{Degree of Freedom}

\newacronym{PI}{PI}{Proportional-Integral}
\newacronym{PID}{PID}{Proportional-Integral-Derivative}
\newacronym{PPO}{PPO}{Proximal Policy Optimization}

\newacronym{DQN}{DQN}{Deep Q-Network}
\newacronym{RLS}{RLS}{Recursive Least-Squares}

\newacronym{TRPO}{TRPO}{Trust Region Policy Optimization}

\newacronym{ReLu}{ReLu}{Rectified Linear Unit}

\newacronym{LTI}{LTI}{Linear Time Invariant}

\newacronym{CPI}{CPI}{Conservative Policy Iteration}

\newacronym{SGD}{SGD}{Stochastic Gradient Descent}

\newacronym{KL}{KL}{Kullback–Leibler}

\newacronym{SAC}{SAC}{Soft Actor-Critic}

\newacronym{ACD}{ACD}{Adaptive Critic Designs}
\newacronym{MDP}{MDP}{Markov Decision Process}

\begin{document}

\maketitle

\begin{abstract}

Fault-tolerant flight control faces challenges, as developing a model-based controller for each unexpected failure is unrealistic, and online learning methods can handle limited system complexity due to their low sample efficiency. In this research, a model-free coupled-dynamics flight controller for a jet aircraft able to withstand multiple failure types is proposed. An offline-trained cascaded \acrlong{SAC} \acrlong{DRL} controller is successful on highly coupled maneuvers, including a coordinated $\boldsymbol{40^\circ}$-bank climbing turn with a normalized Mean Absolute Error of $\boldsymbol{2.64\%}$. The controller is robust to six failure cases, including the rudder jammed at $\boldsymbol{-15^\circ}$, the aileron effectiveness reduced by $\boldsymbol{70\%}$, a structural failure, icing and a backward c.g. shift as the response is stable and the climbing turn is completed successfully. Robustness to biased sensor noise, atmospheric disturbances, and to varying initial flight conditions and reference signal shapes is also demonstrated.

\end{abstract}

\section*{Nomenclature}

{\renewcommand\arraystretch{1.0}
\noindent\begin{longtable*}{@{}l @{\quad=\quad} l@{}}
$\textbf{s},\textbf{a}$ & environment state and actor action vectors\\
$n, m$  & number of environment states and actor actions, respectively \\
$\Tilde{r}(\textbf{s},\textbf{a})$ & instantaneous reward function\\
$t, \Delta t, N$ & discrete time-step subscript, sample time and number of samples \\
$f(\textbf{s},\textbf{a})$ & state transition function \\
$\pi,\pi^*,\pi_\thetaparam$ & policy, optimal policy and parameterized policy approximation\\
$Q^\pi,Q_\mathbf{k}$ & action-state value function (Q-function) and parameterized Q-function approximation \\
$\thetaparam, \mathbf{k}, \bar{\mathbf{k}}$ & policy, Q-function and target Q-function parameter vectors \\
$\gamma$ & discount factor \\
$\eta$ & temperature parameter \\
$\mathcal{H},\Tilde{\mathcal{H}}$ & entropy and target entropy \\
$L(x), J(x)$ & loss and objective functions for variable $x$ \\
$\mathcal{D}, \mathcal{B}$ & memory buffer and minibatch (subset of the memory buffer) \\
$\text{\boldmath$\xi$}$ & noise vector \\
$\mathcal{N}$ & standard normal distribution \\
$ \text{\boldmath$\mu$}, \text{\boldmath$\sigma$}$ & mean and Sample Standard Deviation (SSD) vectors \\
$\lambda$ & learning rate \\
$\tau$ & smoothing factor \\
$\textbf{x}, \textbf{u}$ & aircraft state and control input vectors \\
$ p,q,r, \phi,\theta,\psi$ & roll, pitch and yaw rates, and roll, pitch and yaw angles \\
$ V,\alpha,\beta$ & total airspeed, angle-of-attack and sideslip angle\\
$h, \Delta h$ & altitude and altitude tracking error\\
$\delta_\textrm{e}, \delta_\textrm{a}, \delta_\textrm{r}$ & control surface deflections (elevator, aileron, rudder) \\
$R$ & reference signal superscript\\
$\boldsymbol{\Delta} \textbf{u}, \Delta \theta^R$ & control input and reference pitch angle increments\\
$\textbf{e}, \textbf{c}$ & error and error cost vectors \\
$l$ & number of units per hidden layer\\
$C_L,C_D,C_m$ & lift, drag and pitching moment coefficients \\
$T,A$ & signal period and amplitude \\

\end{longtable*}}

\section{Introduction}

In-flight loss of control was the cause of $61\%$ of commercial flight accident casualties between 2009 and 2018, indicating the need for more fault-tolerant control systems \cite{IATA}. At the same time, the advent of personal air vehicles in dense urban areas calls for the development of autonomous flight controllers that can withstand multiple types of failures. 

Until now, flight control automation techniques have relied on gain-scheduling to switch between parallel linear controllers tuned at specific known operating points \cite{stevens2015aircraft}. Because of their reliance on known plant dynamics, they cannot deal with sudden changes in dynamics such as failures. Model-free adaptive and intelligent control techniques offer the possibility of replacing this inconvenient controller structure with a more general and fault-tolerant approach.\\

\acrfull{RL} is a bio-inspired machine learning framework for intelligent control that can offer fault-tolerance. An \acrshort{RL} agent learns through trial-and-error by interacting with the plant, also known as the environment \cite{sutton2011reinforcement}. In its original form, \acrshort{RL} was a tabular method with discrete action and state spaces, well suited for gaming environments. To combat the curse of dimensionality encountered when making the action and state spaces larger \cite{zbMATH03126094}, function approximators, typically \acrlong{NN}s (\acrshort{NN}s), can be used with the actor-critic agent structure to enable continuous control.

A common approach to actor-critic \rl structures is \acrfull{ADP}. It has been applied to coupled-dynamics flight control for a business jet aircraft \cite{ferrari2004online} and a helicopter \cite{enns2003helicopter} in simulations, yet all based on a known model of plant dynamics and thereby limiting their applicability to a wide range of unforeseen failures. Recent research using online incremental model-learning \acrshort{ADP} techniques (\acrshort{iADP}) has managed to provide adaptive control while eliminating all model dependence. Longitudinal control was proposed first by \cite{zhou2016nonlinear}, and then the introduction of \acrfull{IDHP} demonstrated fault-tolerance on simple failure cases \cite{zhou2018incremental}. Using \acrshort{IDHP} for coupled-dynamics body rate control, \cite{konatala2021reinforcement} showed that that a business jet aircraft could be controlled successfully. It was extended to altitude and attitude control with pre-tuned \acrshort{PID} controllers in \cite{heyer2020online} and \cite{kroezenthesis}, yet as the longitudinal and lateral motions were decoupled severe failures cases where the coupling effects become too dominant were not tested. When the outer-loop \acrshort{PID}s were replaced with \acrshort{IDHP} agents for longitudinal control in \cite{leethesis}, a failure rate of $24\%$ suggested that this method alone does not yet have the reliability and sample efficiency to control online the outer-loop coupled dynamics of 6-\acrshort{DoF} systems.

A novel approach to actor-critic structures has recently been introduced, known as the field of \acrfull{DRL}. Enabling end-to-end offline learning, high-dimensional input spaces such as images were used to surpass human performance on multiple Atari games, as shown by \cite{mnih2013playing} with \acrfull{DQN} for discrete action spaces. \acrshort{DRL} was extended to control applications by \cite{lillicrap2015continuous} with the off-policy \acrfull{DDPG} algorithm thanks to its continuous control abilities. \acrshort{DDPG} flight control applications have been limited to small-scale flying-wings \cite{tsourdos2019developing} and quadcopters \cite{Quinones,hwangbo2017control,koch2019reinforcement}. On-policy \rl algorithm \acrfull{PPO}, proposed by \cite{schulman2017proximal}, has aimed at reducing \acrshort{DDPG}'s learning instability and showed improved policy convergence for a quadcopter UAV in \cite{lopes2018intelligent} and for an unmanned flying-wing aircraft in \cite{bohn2019deep}. State-of-the-art \acrfull{TD3} and \acrfull{SAC} have focused on reducing \acrshort{DDPG}'s critic overestimation bias \cite{fujimoto2018addressing,haarnoja2018soft}. Unlike \acrshort{TD3}, \acrfull{SAC} uses a stochastic policy which was shown to encourage exploration and increase sample efficiency \cite{haarnoja2018soft}. While it is unclear from the authors of \acrshort{TD3} and \acrshort{SAC} which performed best, an independent study found that \acrshort{SAC} outperformed \acrshort{TD3} in terms of sample efficiency on $4$ out of $5$ complex control tasks \cite{SpinningUp2018}. On a quadcopter control task, it recovered from unfavorable initial conditions in \cite{barros2020using}, demonstrating high robustness. \acrshort{SAC} is identified as the most promising \acrshort{DRL} algorithm for this flight control task. Despite \acrshort{DRL}'s ability to learn highly complex tasks, it has not been tested on coupled-dynamics flight control tasks for fixed-wing aircraft. Furthermore, due to its relatively long training time, online learning is expected to be difficult. For this reason, fault-tolerance is mainly achieved through robust control. \acrshort{SAC}'s generalization ability to multiple types of failures is unknown at this point and is to be better understood.

The contribution of this research is to advance state-of-the-art fault tolerant flight control methods by developing a model-free coupled-dynamics flight controller for a jet aircraft that can withstand multiple types of unexpected failures. This research explores the use of \acrshort{DRL} controllers for CS-25 class aircraft generally only employed for small-scale UAVs. For this research, a high-fidelity simulation model of the Cessna Citation 500 will be used, paving the way for future test flights on the PH-LAB research aircraft thanks to its experimental fly-by-wire flight control system.

The foundations of \rl and the \sac algorithm are explained in Section \ref{sec:fund}, followed by a motivation for the controller design in Section \ref{sec:cont}. The results are discussed in Section \ref{sec:ana} and the conclusions are presented in Section \ref{sec:ccl}.

\section{Fundamentals}
\label{sec:fund}

This section introduces the learning framework used in this research, actor-critic \rl and the \rl algorithm at hand.

\subsection{Reinforcement Learning Problem}

The actor-critic \rl framework is composed of an agent that applies action $\textbf{a}_t\in\mathbb{R}^m$ on an environment with state $\textbf{s}_t\in\mathbb{R}^n$ at discrete time step $t$. The next state is determined by a state-transition function unknown to the agent in Eq.~\eqref{eq:nextstate}. It is assumed to have the Markov property, which implies that the environment's history is fully explained by its present state. The agent chooses actions based on the actor's policy, a mapping from the state space $\mathbb{R}^n$ to the action space $\mathbb{R}^m$. If it is stochastic, the action is sampled as shown in Eq.~\eqref{eq:policy}. The environment gives a scalar reward $\Tilde{r}(\textbf{s}_t,\textbf{a}_t)\in\mathbb{R}$ to the agent as a feedback of its action $\textbf{a}_t$ at each time step. The goal is to learn a policy that maximizes the reward over all states.

A critic, defined by an action-state value function, or Q-function, is introduced in Eq.~\eqref{eq:qfunc} to characterize how beneficial it is to be in a given state $\textbf{s}_t$ in terms of future expected reward when taking action $\textbf{a}_t$ and following the policy thereafter. A discount factor, $\gamma$, is used to trade-off immediate and future rewards. The episode comprises of $N$ time steps.

\begin{table}[htbp]
    \centering
    \begin{tabu} to \textwidth { X[0.78c,m]  X[0.76,c,m]  X[1.45,c,m] }
          \begin{equation}
\textbf{s}_{t+1} = f(\textbf{s}_{t},\textbf{a}_{t})
\label{eq:nextstate}
\end{equation} & \begin{equation} \textbf{a}_{t} \sim
\pi\left(\cdot \mid \textbf{s}_t\right) 
\label{eq:policy}
\end{equation} & \begin{equation}
Q^\pi(\textbf{s}_{t},\textbf{a}_{t})=\underset{\textbf{a}_{t+i} \sim \pi}{\mathbb{E}}\left[\sum_{i=0}^{N} \gamma^{i} \Tilde{r}(\textbf{s}_{t+i},\textbf{a}_{t+i})\mid \textbf{s}_t, \textbf{a}_t\right]
\label{eq:qfunc}
\end{equation} 
    \end{tabu}
\end{table}

\vspace{-1cm}

\subsection{Soft Actor-Critic Algorithm}

\acrfull{SAC} is a novel off-policy \acrshort{DRL} algorithm and an extension of the \acrshort{DDPG} algorithm that aims at optimizing a stochastic policy \cite{haarnoja2018soft}. Unlike \acrshort{DDPG}'s deterministic policy, a stochastic policy ensures better exploration and was found to be applicable to broader types of control tasks \cite{hwangbo2017control}. At evaluation time, the mean of the policy distribution is selected to make actions deterministic and ensure consistent performance.

\subsubsection{Entropy}

\acrshort{SAC} adds an entropy term to the standard optimal policy expression in Eq.~\eqref{eq:entropy-obj-pol}, which is a measure of the randomness in its probability distribution. Policy distributions more spread over the action space have higher entropy, which can be measured with the log-likelihood according to Eq.~\eqref{eq:entropy}. The entropy is traded-off against future rewards with the temperature parameter $\eta$\footnote[2]{The symbol $\eta$ is used as temperature parameter instead of $\alpha$ in \cite{haarnoja2018soft}, the original \sac paper, to avoid confusion with the angle-of-attack.}.

\begin{equation}
    \pi^*= \arg\underset{\pi}{\max}\underset{\textbf{a}_{t+i} \sim \pi}{\mathbb{E}} \left[ \sum_{i=0}^{N} \gamma^{i}\Big(\Tilde{r}(\textbf{s}_{t+i},\textbf{a}_{t+i}) + \eta \mathcal{H}\big(\pi\left(\cdot \mid \textbf{s}_{t+i}\right)\big)\Big) \right] \quad \forall\: \textbf{s}_t \in \mathbb{R}^m
    \label{eq:entropy-obj-pol}
\end{equation}

 \begin{equation}
\mathcal{H}(\pi\left(\cdot \mid \textbf{s}_t\right))=\underset{\textbf{a}^\prime \sim \pi}{\mathbb{E}}[-\log \pi\left(\textbf{a}^\prime \mid \textbf{s}_t\right)]
\label{eq:entropy}
\end{equation}

In other words, this objective favors the most random policy that still achieves a high return. This creates an inherent exploration mechanism that also prevents premature convergence to local optima. Multiple control strategies that achieve a near-optimal reward are captured, allowing for more robustness to disturbances.

\subsubsection{Critic}

The Q-function critic is modeled as a feed-forward \acrfull{DNN} with parameter vector $\textbf{k}$. The standard Bellman equation is modified with the expression of the entropy found in Eq.~\eqref{eq:entropy} to obtain a recursive expression of the soft Q-function in Eq.~\eqref{eq:qfunc-entropy}.

\begin{equation}
Q_\textbf{k}(\textbf{s}_t, \textbf{a}_t) =\underset{\textbf{a}_t, \textbf{a}_{t+1} \sim \pi_\thetaparam}{\mathbb{E}}\left[\Tilde{r}(\textbf{s}_t,\textbf{a}_t)+\gamma\left(Q_\textbf{k}(\textbf{s}_{t+1}, \textbf{a}_{t+1})-\eta \log \pi_{\thetaparam}(\textbf{a}_{t+1} \mid \textbf{s}_{t+1})\right)\right].
\label{eq:qfunc-entropy}
\end{equation}

Given that \sac is an off-policy algorithm, transition samples $(\textbf{s}_t$, $\textbf{a}_t$, $r_t$, $\textbf{s}_{t+1})$ can be collected in a memory buffer $\mathcal{D}$ and reused at a later stage. This can help ensure training samples are more independent and identically distributed, an assumption of the update method of \rl algorithms. By sampling a minibatch $\mathcal{B}$ from the memory buffer, a minibatch gradient update can be performed instead of the computationally inefficient and noisy stochastic gradient update of on-policy algorithms.

To increase learning stability, a target network with parameter $\Bar{\textbf{k}}$ for the Q-function is introduced to make the gradient update follow a more constant direction. From time to time, the target network is synchronized with the current value network with an exponentially weighted moving average as a soft update mechanism, such that the target network is a delayed version of the current value network regulated by smoothing factor $\tau$. The target network is used in the mean squared Bellman error for the Q-function loss function shown in Eq.~\eqref{eq:loss-q}. In an effort to reduce \acrshort{DDPG}'s overestimation bias of the Q-function, \sac makes use of the double Q-function trick by learning two approximators and using a pessimistic bound over the two. Transition samples are contained in minibatch $\mathcal{B}$ but because \sac is off-policy, fresh actions $\textbf{a}_{t+1}$ can be sampled from the current policy to compute the Q-function targets.

\begin{equation}
    L_Q\left(\textbf{k}_{i},\mathcal{B}\right)=\underset{\begin{subarray}{c}(\textbf{s}_t,\textbf{a}_t,\textbf{s}_{t+1}) \sim \mathcal{B}\\\textbf{a}_{t+1} \sim \pi_\thetaparam\end{subarray}}{\mathbb{E}}\left[\left(Q_{\textbf{k}_{i}}(\textbf{s}_t, \textbf{a}_t)-\left(\Tilde{r}(\textbf{s}_t,\textbf{a}_t)+\gamma\left(\min _{i=1,2} Q_{\Bar{\textbf{k}}_{i}}(\textbf{s}_{t+1}, \textbf{a}_{t+1})-\eta \log \pi_{\thetaparam}(\textbf{a}_{t+1} \mid \textbf{s}_{t+1})\right)\right)\right)^{2}\right]
    \label{eq:loss-q}
\end{equation}

\subsubsection{Policy}

The policy, or actor, is modeled as an $m$-dimensional multivariate Gaussian distribution with a diagonal covariance matrix. Its actions are passed to a $\tanh$ squashing function to ensure they are defined on a finite bound. The mean vector $\text{\boldmath$\mu_\theta$}$ and the covariance matrix, or, in this case, vector $\text{\boldmath$\sigma_\theta^2$}$, are estimated for each state by a \acrshort{DNN} with parameter vector $\thetaparam$.

Unlike \acrshort{DDPG}'s deterministic policy, no target policy is needed as the policy's stochasticity has a smoothing effect. The stochasticity also means that the policy objective in Eq.~\eqref{eq:entropy-obj-pol} depends on the expectation over actions and is therefore non-differentiable. A reparameterization trick is proposed by the \sac authors in \cite{haarnoja2018soft} using the known mean and standard deviation of the stochastic policy along with independent noise vector $\text{\boldmath$\xi$}$, and applying the squashing function, as shown in Eq.~\eqref{eq:reparam}. A policy objective with an expectation over noise instead of actions and making use of the Q-function as an approximation of expected future rewards and entropy is introduced in Eq.~\eqref{eq:entropy-obj-pol2}.

\begin{equation}
\Tilde{\textbf{a}}_\thetaparam(\textbf{s}_t,\text{\boldmath$\xi$}) = \tanh\left( \text{\boldmath$\mu_\theta$}(\textbf{s}_t) + \text{\boldmath$\sigma_\theta$}(\textbf{s}_t) \odot \text{\boldmath$\xi$}\right), \quad \text{\boldmath$\xi$} \sim \mathcal{N}(\vec{\textbf{0}} ,\vec{\textbf{1}})
    \label{eq:reparam}
\end{equation}

\begin{equation}
    J_\pi\left(\thetaparam,\mathcal{B}\right) = \underset{\begin{subarray}{c}\textbf{s}_t \sim \mathcal{B}\\\text{\boldmath$\xi$} \sim \mathcal{N}\end{subarray}}{\mathbb{E}} \left[ \min _{i=1,2} Q_{\textbf{k}_{i}}(\textbf{s}_t, \Tilde{\textbf{a}}_\thetaparam(\textbf{s}_t,\text{\boldmath$\xi$}))-\eta \log \pi_{\thetaparam}(\Tilde{\textbf{a}}_\thetaparam(\textbf{s}_t,\text{\boldmath$\xi$}) \mid \textbf{s}_t)\right]
    \label{eq:entropy-obj-pol2}
\end{equation}

\subsubsection{Automatic Temperature Adjustment}

\sac can be unstable with respect to temperature parameter $\eta$, so the latter was proposed to be controlled automatically in \cite{haarnoja2018soft2}. Optimal entropy is not constant throughout training as the need for exploration is expected to decrease with increasing training steps. A loss function $L\left(\eta\right)$ is introduced in Eq.~\eqref{eq:auto-entropy} to dynamically find the lowest temperature that still ensures a certain minimum target entropy $\bar{\mathcal{H}}$ while maximizing the return. A good empirical value for the target entropy is found to be connected to the action space dimension with $\log \bar{\mathcal{H}}= -m$ \cite{haarnoja2018soft2}.

\begin{equation}
    L\left(\eta\right) = \underset{\begin{subarray}{c}\textbf{s}_t \sim \mathcal{B}\\\textbf{a}_t \sim \pi_\thetaparam\end{subarray}}{\mathbb{E}} \left[ -\eta \log \pi_{\thetaparam}(\textbf{a}_t \mid \textbf{s}_t)  - \eta \bar{\mathcal{H}} \right]
    \label{eq:auto-entropy}
\end{equation}

\subsubsection{Overview}

\begin{figure}[H]
    \centering
    \includegraphics[width=1.05\textwidth]{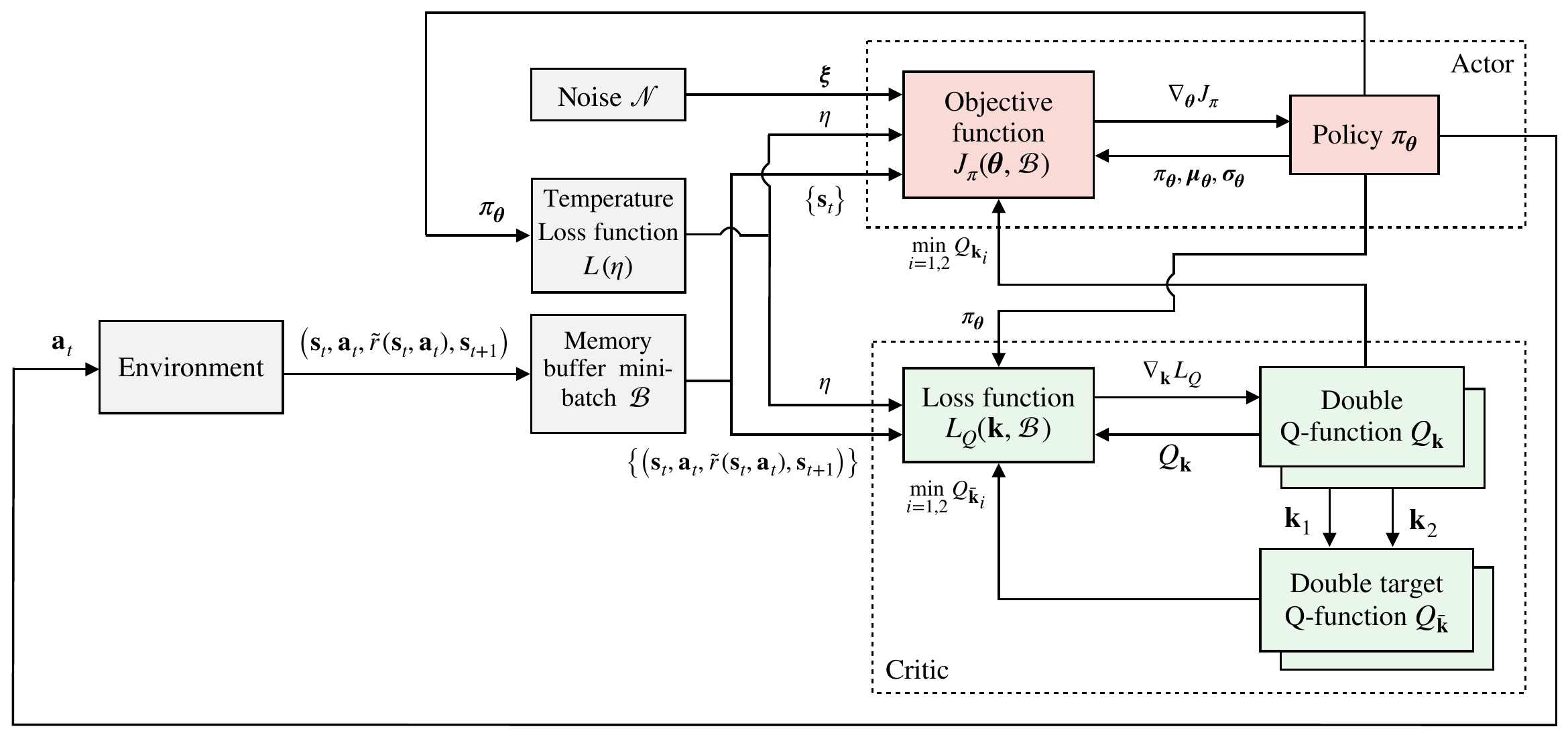}
    \caption{\sac framework.}
    \label{fig:framework}
\end{figure}

An overview of the \sac framework is shown in Fig.~\ref{fig:framework}. With the critic loss and policy objective functions, a pseudocode can be constructed as shown in Algorithm \ref{algo:sac}.

\begin{algorithm}[H]
\SetAlgoLined
 Initialize $\thetaparam$, $\textbf{k}_1$, $\textbf{k}_2$ parameters for policy $\pi_\thetaparam$ and double Q-function $Q_{\textbf{k}_{1,2}}$\;
 Set target parameters $\Bar{\textbf{k}}_1 \leftarrow \textbf{k}_1$ and $\Bar{\textbf{k}}_2 \leftarrow \textbf{k}_2$\;
 
 Initialize empty memory buffer $\mathcal{D}$, minibatch $\mathcal{B}$, learning rate $\lambda$ and smoothing factor $\tau$ \;
 
  Observe initial state $\textbf{s}_0$\;
 
 \For{\textrm{each time step} $t$}{

 Sample action $\textbf{a}_t \sim \pi_\thetaparam\left(\cdot \mid \textbf{s}_{t}\right)$ \;
 Observe next state and reward $\textbf{s}_{t+1}=f(\textbf{s}_{t},\textbf{a}_{t})$, $\Tilde{r}(\textbf{s}_t,\textbf{a}_t)$ \; 
 
Store transition sample $\left(\textbf{s}_t, \textbf{a}_t, \Tilde{r}(\textbf{s}_t,\textbf{a}_t), \textbf{s}_{t+1}\right)$ in $\mathcal{D}$ \;

 Sample a minibatch of transition samples $\mathcal{B}=\left\{\left(\textbf{s}_t, \textbf{a}_t, \Tilde{r}(\textbf{s}_t,\textbf{a}_t), \textbf{s}_{t+1}\right)\right\}$ from $\mathcal{D}$ \;
 
 Update Q-function parameters: $\textbf{k}_i \leftarrow \textbf{k}_i - \lambda \nabla_{\textbf{k}_{i}} \frac{1}{|\mathcal{B}|}\sum L_Q(\textbf{k}_i,\mathcal{B})$ for $i={1,2}$\;
 
 Update policy parameter: $\thetaparam \leftarrow \thetaparam + \lambda \nabla_{\thetaparam} \frac{1}{|\mathcal{B}|}\sum J_\pi(\thetaparam,\mathcal{B})  $

 Update the temperature hyperparameter: $\eta \leftarrow \eta - \lambda \nabla_{\eta} L\left(\eta\right)$ \;
  \vspace{1mm}
 Update target Q-function parameters: $ \Bar{\textbf{k}}_i \leftarrow (1-\tau) \textbf{k}_i + \tau \Bar{\textbf{k}}_i$ for $i=1,2$\;

}
\caption{\acrshort{SAC}. Adapted from \cite{haarnoja2018soft2}.}
 \label{algo:sac}
\end{algorithm}

\section{Controller Design}
\label{sec:cont}

With the \sac framework presented above, its integration with the flight controller is discussed in this section.

\subsection{High-Fidelity Cessna Citation 500 Model}

\begin{figure}[H]
\centering
\includegraphics[width=.5\textwidth]{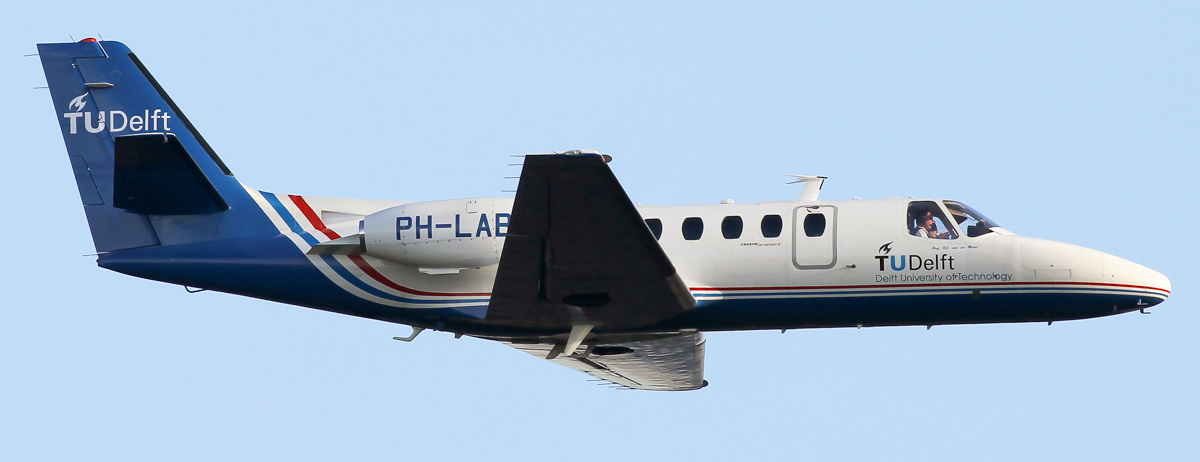}
\caption[Cessna Citation PH-LAB research aircraft]{Cessna Citation PH-LAB research aircraft.\footnotemark[1]}
\label{fig:cit}
\end{figure}

The system to be controlled in this application is a high-fidelity non-linear simulation model of the Cessna Citation 500 business jet aircraft. It was built with the \acrfull{DASMAT} based on flight data recorded on the PH-LAB research aircraft shown in Fig.~\ref{fig:cit}. The model was validated in \cite{van2018identification}. It is expected that the controller proposed in this research will be flight-tested on the PH-LAB at a later stage.

\footnotetext[1]{Image from C. v. Grinsven (with permission).}

The full coupled-dynamics of the aircraft are to be controlled with a refresh rate of $100$Hz. For this research, actuator dynamics are modeled with a low-pass filter and saturation limits, while ideal sensors are assumed. A yaw damper is already present on the aircraft. The aircraft is kept in a clean configuration for this simulation. The aircraft state and control input vectors are given in Eqs.~\eqref{eq:state-cit} and \eqref{eq:controlinput-cit}, respectively. At the beginning of each simulation, the aircraft is untrimmed with null initial control inputs.

\begin{multicols}{2}
\noindent
\begin{equation}
    \textbf{x} = \left[ p,q,r,V,\alpha,\beta,\theta,\phi,\psi,h \right]^\intercal
    \label{eq:state-cit}
\end{equation}
\columnbreak
\noindent
\begin{equation}
    \textbf{u} = \left[ \delta_\textrm{e}, \delta_\textrm{a}, \delta_\textrm{r}\right]^\intercal
    \label{eq:controlinput-cit}
\end{equation}
\end{multicols}
\vspace{-0.7cm}

\subsection{Interfacing}

In reinforcement learning, the flight controller, the plant and the control input are known as the agent, the environment and the action, respectively.

A flight controller for automatic altitude and attitude tracking is to be built as part of this research so that it can be interfaced with existing navigation algorithms reviewed in \cite{delahaye2014mathematical}. To ensure compatibility with the experimental fly-by-wire system of the PH-LAB in future flight tests, the controller developed in this research will only command control surfaces while the airspeed is controlled with an independent \acrshort{PID} auto-throttle. A flight control task for altitude, roll and sideslip angles tracking is proposed. Because of the difference in their dynamics, a more stable learning is expected by having the altitude and attitude controlled by two separate, cascaded controllers.

\subsubsection{Attitude Control}

An inner-loop \sac agent will track reference signals for the pitch, roll and sideslip angles, referred to with the $R$ superscript. A reward function based on the clipped L1 norm of the error vector is proposed in Eq.~\eqref{eq:reward-agent1}. A cost vector $\textbf{c}$, determined by trial and error, is associated with the tracked states in Eq.~\eqref{eq:cost}, where the sideslip angle is attributed a higher cost due to its generally low magnitude.

\begin{multicols}{2}
\noindent
\begin{equation}
    \textbf{e}^\text{att} = \left[\beta^R-\beta,\theta^R-\theta,\phi^R-\phi \right]^\intercal
    \label{eq:error}
\end{equation}
\columnbreak
\noindent
\begin{equation}
\textbf{c}^\text{att} = \frac{6}{\pi}\left[4,1,1\right]^\intercal
\label{eq:cost}
\end{equation}
\end{multicols}
\vspace{-0.8cm}
\begin{equation}
 \Tilde{r}(\textbf{e}^\text{att}) = -\frac{1}{3}\left\|\operatorname{clip}\left[ \textbf{c}^\text{att}\odot\textbf{e}^\text{att},\vec{\textbf{-1}},\vec{\textbf{0}}\right]\right\|_1
    \label{eq:reward-agent1}
\end{equation}

Initial tests found that the aircraft control input was noisy when corresponding directly to the agent action. To smooth out the control input, the agent is set to command the control input increment $\boldsymbol{\Delta} \textbf{u}$ in Eq.~\eqref{eq:control-input}. Because the $\tanh$ function squashes the action vector $\textbf{a}^\text{att}$ to $\left[-1,1\right]^3$, it is mapped to the physical range of the control input increment in Eq.~\eqref{eq:actions-agent1}, chosen as a hundredth of actuator limits to prevent sharp variations.

\begin{multicols}{2}
\noindent
\begin{equation}
   \textbf{u}_{t} = \textbf{u}_{t-1} + \boldsymbol{\Delta} \textbf{u}_t
   \label{eq:control-input}
\end{equation}
\columnbreak
\begin{equation}
    \boldsymbol{\Delta} \textbf{u} = \boldsymbol{\Delta} \textbf{u}_\text{min}+(\textbf{a}^\text{att}+1)\frac{\boldsymbol{\Delta}\textbf{u}_\text{max}-\boldsymbol{\Delta}\textbf{u}_\text{min}}{2}
       \label{eq:actions-agent1}
\end{equation}
\noindent
\end{multicols}
\vspace{-1mm}

A trade-off has to be made between making the environment state smaller to speed up learning and giving enough information to allow the agent to make informed decisions. As the environment is assumed to have the Markov property, only information of the current time step is needed to explain its state. The environment state is decided to contain the weighted error vector to ensure a satisfactory steady-state response, the three body rates to improve the transient response, and the current control input since the agent only controls its increment. The environment state is given in Eq.~\eqref{eq:state-agent1}. 

\begin{equation}
   \textbf{s}^\text{att} = \left[\left(\textbf{c}^\text{att}\odot \textbf{e}^\text{att}\right)^\intercal,\textbf{u}^\intercal,p,q,r\right]^\intercal
   \label{eq:state-agent1}
\end{equation}

\subsubsection{Altitude Control}

An outer-loop \sac agent is tasked with providing a reference pitch angle to track an altitude signal. The error and cost vectors are defined in Eqs.~\eqref{eq:error2} and \eqref{eq:cost-alt}, respectively. Similarly to the inner-loop agent, the reward function is defined as the absolute clipped weighted error in Eq.~\eqref{eq:reward-agent2}. The cost was found empirically so that, despite the clipping, differences between large errors are perceived by the agent, while still incentivizing for a low steady-state error.

\begin{table}[htbp]
    \centering
    \begin{tabu} to \textwidth { X[0.78c,m]  X[0.5,c,m]  X[1.5,c,m] }
          \begin{equation}
    \mathbf{e}^\text{alt} = \left[h^R-h\right] = \left[\Delta h\right] 
    \label{eq:error2}
\end{equation} & \begin{equation}
    \textbf{c}^\text{alt} = \left[\frac{1}{240}\right]
    \label{eq:cost-alt}
\end{equation} & \begin{equation}
 \Tilde{r}(\textbf{e}^\text{alt}) = -\left\|\operatorname{clip}\left[ \textbf{c}^\text{alt}\odot\textbf{e}^\text{alt},\vec{\textbf{-1}},\vec{\textbf{0}}\right]\right\|_1
    \label{eq:reward-agent2}
\end{equation}
    \end{tabu}
\end{table}

For this agent too, the control policy is smoothed by controlling the reference pitch angle increment instead of its value in \eqref{eq:ref-pitch}. The agent action $\textbf{a}^\text{alt}$ defined on $\left[-1,1\right]$ is mapped to the pitch angle increment in Eq.~\eqref{eq:actions-agent2} such that the corresponding pitch rate does not exceed $10\text{deg}\:\text{s}^{-1}$.

\begin{multicols}{2}
\noindent
\begin{equation}
   \theta_{t}^R = \theta_{t-1}^R + \Delta \theta^R_t
   \label{eq:ref-pitch}
\end{equation}
\columnbreak
\begin{equation}
    \Delta \theta^R = \Delta \theta^R_\text{min} + (\textbf{a}^\text{alt}+1)\frac{\Delta \theta^R_\text{max} -\Delta \theta^R_\text{min}}{2}
       \label{eq:actions-agent2}
\end{equation}
\noindent
\end{multicols}

The state environment in Eq.~\eqref{eq:state-agent2} is reduced because the agent only has to learn the kinematic relationship between altitude and pitch angle. No knowledge of lateral states is needed since the inner-loop controller is already fully coupled. 

\begin{equation}
   \textbf{s}^\text{alt} = \left[ \textbf{c}^\text{alt}\odot\textbf{e}^\text{alt},\theta^R\right]^\intercal
   \label{eq:state-agent2}
\end{equation}

\begin{figure}[H]
    \centering
    \includegraphics[width=1.05\textwidth]{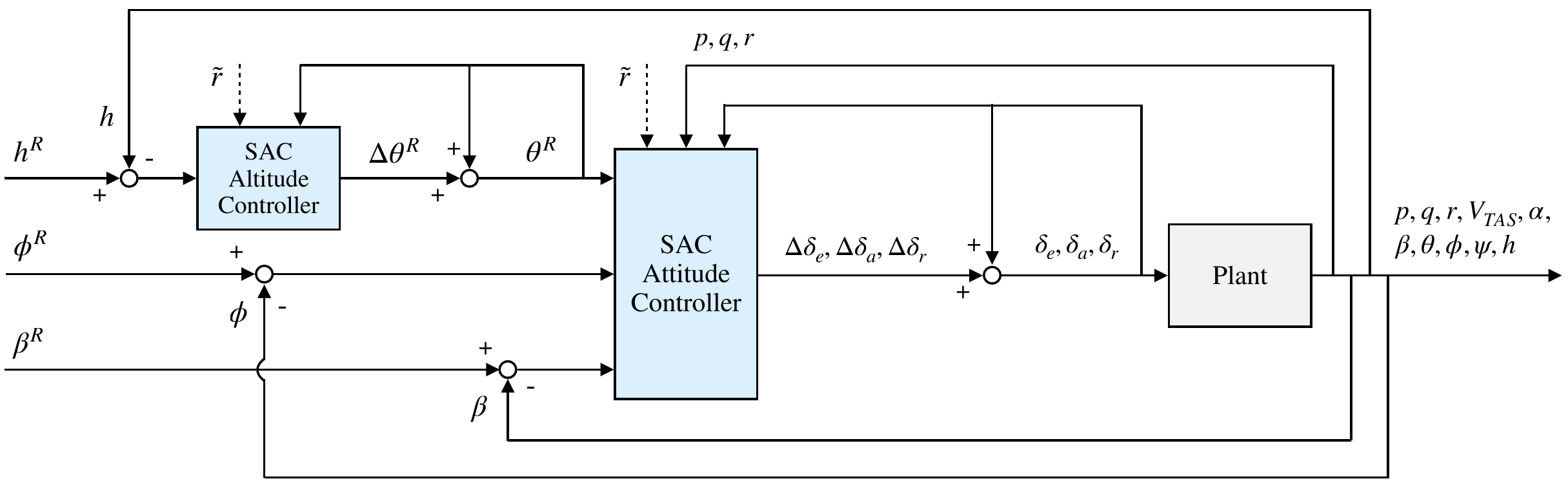}
    \caption{Cascaded controller structure for altitude and attitude control. The \sac controllers observe the weighted errors, untracked states, and current control input and also receive a reward from the environment.}
    \label{fig:control-diagram}
\end{figure}

A diagram of the cascaded controller structure is shown in Fig.~\ref{fig:control-diagram}. Feedback signals of plant state variables, current pitch reference angle and control surface deflections are from the previous time step.

\subsubsection{Hyperparameters}

 Choosing suitable hyperparameters for the \sac controllers and \acrshort{DNN}s can significantly improve sample efficiency. A short selective-search hyperparameter optimization centered on default values was performed on the learning rate, number of hidden units, memory buffer and minibatch sizes, all known to be influential hyperparameters. In particular, it was found that the altitude controller, with smaller state and action spaces, performed better with fewer hidden units and a higher average learning rate. Several default hyperparameters are used from the original \sac implementation \cite{haarnoja2018soft2}. An overview of chosen hyperparameters is presented in Table.~\ref{tab:hyper}. Additionally, network initialization is executed following the Xavier method \cite{glorot2010understanding}, and the gradient-descent employs the Adam optimizer, recognized for its superiority in terms of learning stability \cite{ruder2016overview}.\\

\begin{table}
\centering
\caption{Cascaded \sac controller hyperparameters. Default values from \cite{haarnoja2018soft2}.}
\label{tab:hyper}
\begin{tabular}{lll} 
\toprule
\textbf{Hyperparameter} & \textbf{Altitude Controller} & \textbf{Attitude Controller} (if different)                                                                       \\[5pt]

\cmidrule(r){1-3}
Learning rate $\lambda$                       &$3\cdot 10^{-4}$&Linearly decreasing from $4\cdot 10^{-4}$ to $0$ \\
Hidden units $l$x$l$                & $32$x$32$                 & $64$x$64$                                                                         \\
Entropy target~$\log \Bar{\mathcal{H}}$                  & $-m=-1$ (default)       &$-m=-3$ (default)                                                                              \\
Discount factor~$\gamma$                        & $0.99$ (default)                                                                                           \\
Network activation                     & ReLu (default)                                                                    \\
Memory buffer size $|\mathcal{D}|$                                      & $5\cdot 10^4$   \\
Minibatch size $|\mathcal{B}|$                                    & $256$ (default)                                                                                       \\
Smoothing factor $\tau$                   & $0.995$ (default)                                                                               \\   
\bottomrule
\end{tabular}
\end{table}

\tikzset{%
  every neuron/.style={
    circle,
    draw,
    minimum size=0.5cm
  },
  neuron missing/.style={
    draw=none, 
    scale=1,
    text height=0.333cm,
    execute at begin node=\color{black}$\vdots$
  },
}

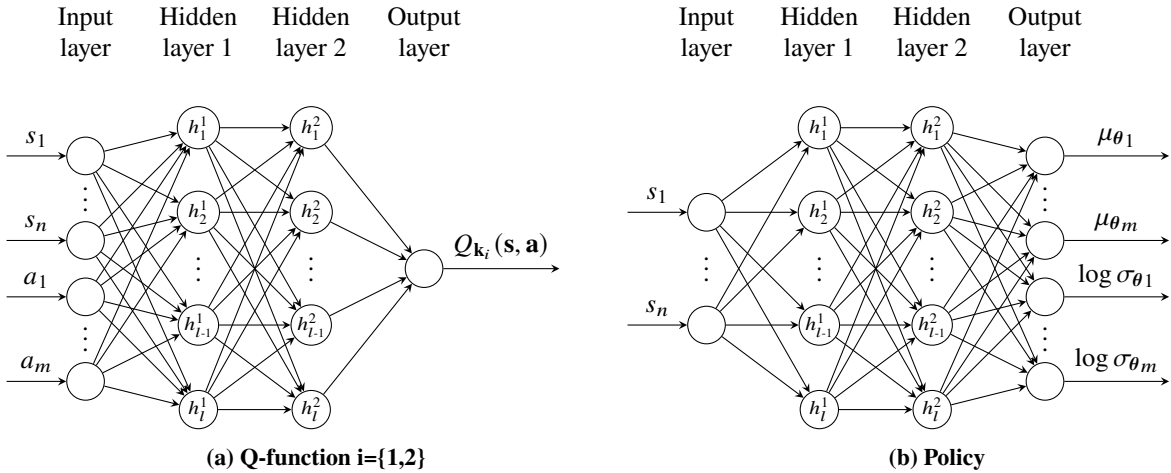
\begin{figure}[H]
  \centering
  \begin{minipage}[b]{0.5\textwidth}
    
\begin{tikzpicture}[x=1.5cm, y=1.5cm, >=stealth]

\node [every neuron/.try, neuron 1/.try] (input-1) at (0,1.25) {};
\node [every neuron/.try, neuron missing/.try] (input-missing) at (0,0.87) {};
\node [every neuron/.try, neuron 2/.try] (input-2) at (0,0.5) {};
\node [every neuron/.try, neuron 3/.try] (input-3) at (0,-0) {};
\node [every neuron/.try, neuron missing/.try] (input-missing) at (0,-0.37) {};
\node [every neuron/.try, neuron 4/.try] (input-4) at (0,-0.75) {};

\foreach \m/\l [count=\y] in {1,2}
  \node [inner sep=0.05cm, every neuron/.try, neuron \m/.try] (hidden1-\m) at (1,2.25-0.75*\y) {\scriptsize$h^1_\m$};
\node [inner sep=0.05cm,every neuron/.try, neuron missing/.try ] (hidden1-missing) at (1,0.25) {};
\node [inner sep=0.015cm, every neuron/.try, neuron p-1/.try] (hidden1-p-1) at (1,0.5-0.75*1) {\scriptsize$h^1_{\tiny \scalebox{.95}{\ensuremath{l}-1}}$};
\node [inner sep=0.03cm, every neuron/.try, neuron p/.try] (hidden1-p) at (1,0.5-0.75*2) {\scriptsize$h^1_{l}$};

\foreach \m/\l [count=\y] in {1,2}
  \node [inner sep=0.05cm, every neuron/.try, neuron \m/.try] (hidden2-\m) at (2,2.25-0.75*\y) {\scriptsize$h^2_\m$};
\node [inner sep=0.05cm,every neuron/.try, neuron missing/.try ] (hidden2-missing) at (2,0.25) {};
\node [inner sep=0.015cm, every neuron/.try, neuron p-1/.try] (hidden2-p-1) at (2,0.5-0.75*1) {\scriptsize$h^2_{\tiny \scalebox{.95}{\ensuremath{l}-1}}$};
\node [inner sep=0.03cm, every neuron/.try, neuron p/.try] (hidden2-p) at (2,0.5-0.75*2) {\scriptsize$h^2_{l}$};

\foreach \m [count=\y] in {1}
  \node [every neuron/.try, neuron \m/.try ] (output-\m) at (3,0.75-0.5*\y) {};

\draw [<-] (input-1) -- ++(-0.7,0)
node [above, midway] {$s_1$};
\draw [<-] (input-2) -- ++(-0.7,0)
node [above, midway] {$s_n$};
\draw [<-] (input-3) -- ++(-0.7,0)
node [above, midway] {$a_1$};
\draw [<-] (input-4) -- ++(-0.7,0)
node [above, midway] {$a_m$};

\draw [->] (output-1) -- ++(1.2,0)
node [above, midway] {$Q_{\textbf{k}_{i}}(\textbf{s},\textbf{a})$};

\foreach \i in {1,...,4}
  \foreach \j in {1,2,p-1,p}
    \draw [->] (input-\i) -- (hidden1-\j);
    
\foreach \i in {1,2,p-1,p}
  \foreach \j in {1,2,p-1,p}
    \draw [->] (hidden1-\i) -- (hidden2-\j);

\foreach \i in {1,2,p-1,p}
    \draw [->] (hidden2-\i) -- (output-1);

\foreach \l [count=\x from 0] in {Input \\ layer, Hidden \\ layer 1, Hidden \\ layer 2, Output \\ layer}
  \node [align=center, above] at (\x,2) {\l};

\end{tikzpicture}
    \subcaption{Q-function i=\{1,2\}}
	\label{fig:q-func-net}
  \end{minipage}%
  \hfill
  \begin{minipage}[b]{0.5\textwidth}

\begin{tikzpicture}[x=1.5cm, y=1.5cm, >=stealth]

\foreach \m/\l [count=\y] in {1,missing,2}
  \node [every neuron/.try, neuron \m/.try] (input-\m) at (0,1.75-0.5*\y) {};

\foreach \m/\l [count=\y] in {1,2}
  \node [inner sep=0.05cm, every neuron/.try, neuron \m/.try] (hidden1-\m) at (1,2.75-0.75*\y) {\scriptsize$h^1_\m$};
\node [inner sep=0.05cm,every neuron/.try, neuron missing/.try ] (hidden1-missing) at (1,0.75) {};
\node [inner sep=0.015cm, every neuron/.try, neuron p-1/.try] (hidden1-p-1) at (1,1-0.75*1) {\scriptsize$h^1_{\tiny \scalebox{.95}{\ensuremath{l}-1}}$};
\node [inner sep=0.03cm, every neuron/.try, neuron p/.try] (hidden1-p) at (1,1-0.75*2) {\scriptsize$h^1_{l}$};

\foreach \m/\l [count=\y] in {1,2}
  \node [inner sep=0.05cm, every neuron/.try, neuron \m/.try] (hidden2-\m) at (2,2.75-0.75*\y) {\scriptsize$h^2_\m$};
\node [inner sep=0.05cm,every neuron/.try, neuron missing/.try ] (hidden2-missing) at (2,0.75) {};

\node [inner sep=0.015cm, every neuron/.try, neuron p-1/.try] (hidden2-p-1) at (2,1-0.75*1) {\scriptsize$h^2_{\tiny \scalebox{.95}{\ensuremath{l}-1}}$};
\node [inner sep=0.03cm, every neuron/.try, neuron p/.try] (hidden2-p) at (2,1-0.75*2) {\scriptsize$h^2_{l}$};

\node [every neuron/.try, neuron 1/.try] (output-1) at (3,1.75) {};
\node [every neuron/.try, neuron missing/.try] (input-missing) at (3,1.37) {};
\node [every neuron/.try, neuron 2/.try] (output-2) at (3,1) {};
\node [every neuron/.try, neuron 3/.try] (output-3) at (3,0.5) {};
\node [every neuron/.try, neuron missing/.try] (input-missing) at (3,0.13) {};
\node [every neuron/.try, neuron 4/.try] (output-4) at (3,-0.25) {};

\draw [<-] (input-1) -- ++(-0.7,0)
node [above, midway] {$s_1$};
\draw [<-] (input-2) -- ++(-0.7,0)
node [above, midway] {$s_n$};

\draw [->] (output-1) -- ++(1.1,0)
node [above, midway] {${\mu_\thetaparam}_1$};
\draw [->] (output-2) -- ++(1.1,0)
node [above, midway] {${\mu_\thetaparam}_m$};

\draw [->] (output-3) -- ++(1.1,0)
node [above, midway] {$\log{\sigma_\thetaparam}_1$};
\draw [->] (output-4) -- ++(1.1,0)
node [above, midway] {$\log{\sigma_\thetaparam}_m$};

\foreach \i in {1,...,2}
  \foreach \j in {1,2,p-1,p}
    \draw [->] (input-\i) -- (hidden1-\j);
    
\foreach \i in {1,2,p-1,p}
  \foreach \j in {1,2,p-1,p}
    \draw [->] (hidden1-\i) -- (hidden2-\j);

\foreach \i in {1,2,p-1,p}
  \foreach \j in {1,...,4}
    \draw [->] (hidden2-\i) -- (output-\j);

\foreach \l [count=\x from 0] in {Input \\ layer, Hidden \\ layer 1, Hidden \\ layer 2, Output \\ layer}
  \node [align=center, above] at (\x,2.5) {\l};

\end{tikzpicture}
	\subcaption{Policy}
	\label{fig:pol-net}
  \end{minipage}

  \caption{Network topology of SAC controllers. The altitude controller has $\boldsymbol{n=2}$, $\boldsymbol{m=1}$ and $\boldsymbol{l=32}$ while the attitude controller has $\boldsymbol{n=9}$, $\boldsymbol{m=3}$ and $\boldsymbol{l=64}$. Subscript $\boldsymbol{j}$ of $\boldsymbol{s_j}$ represents the $\boldsymbol{j}$-th element of vector \textbf{s}.}
  \label{fig:topo}
\end{figure}

The network topology for the controllers is visualized in Fig.~\ref{fig:topo}. The double Q-function corresponds to two independent networks with the structure shown in Fig.~\ref{fig:q-func-net}. The policy's multivariate Gaussian distribution is defined according to the mean and log-standard deviation vectors from the network output in Fig.~\ref{fig:pol-net}. Using log-standard deviations allows the network to estimate the parameter on $\mathbb{R}$ and exponentiation is used to sample actions from the policy according to Eq.~\eqref{eq:reparam}. At evaluation time, actions are made deterministic for consistent performance by setting the noise vector in Eq.~\eqref{eq:reparam} to zero.

\begin{figure}[H]
  \centering
\begin{tikzpicture}[>=stealth,
init/.style={
  draw,
  circle,
  inner sep=2pt,
  font=\huge,
  join = by ->
},
squa/.style={
  draw,
  inner sep=4pt,
  font=\normalsize,
  join = by ->
},
start chain=2,node distance=13mm
]
\node[on chain=2] at (1cm,0cm) (k2)
  {$w_2$};
\draw [<-] (k2) -- ++(-1,0)
node [above, left] {$x_2$};

\node[on chain=2,init] at (0.7,0cm)  (sigma) 
  {$\displaystyle\Sigma$};
  
\node[on chain=2,squa,label=above:{\parbox{2cm}}] at (2.2,0cm)   
  {\begin{tabular}[c]{@{}c@{}}Norm.\end{tabular}};
\node[on chain=2,squa,label=above:{\parbox{2cm}{\centering Activation \\ function}}] at (4.05,0cm)   
  {\begin{tabular}[c]{@{}c@{}}ReLu\end{tabular}};
\node[on chain=2,label=above:Output,join=by ->] at (6.0,0cm) 
  {$y$};
  
\begin{scope}[start chain=1]
\node[on chain=1,label=above:Weights,join=by ->] at (1,0.7cm) 
  (k1) {$w_1$};
\draw [<-] (k1) -- ++(-1,0)
node [above, left] {$x_1$};
\end{scope}
\begin{scope}[start chain=3]
\node[on chain=3] at (1cm,-0.85cm) 
  (k3) {$w_i$};
\draw [<-] (k3) -- ++(-1,0)
node [above, left] {$x_i$};
\end{scope}
\node [every neuron/.try, neuron missing/.try] (input-missing) at (-0.3,-0.4) {};
\node [every neuron/.try, neuron missing/.try] (input-missing) at (1,-0.4) {};
\node[label=above:\parbox{0.4cm}{\centering Bias \\ $b$}] at ([yshift=1cm]sigma) (b) {};

\node[label=above:\parbox{0.5cm}{\centering Inputs}] at (-0.5,0.85) (inputs) {};

\draw [->] (k1) -- (sigma);
\draw [->] (k3) -- (sigma);
\draw [->] (b) -- (sigma);

\end{tikzpicture}
 \caption{Hidden unit $\boldsymbol{h}$ with input vector $\boldsymbol{x}$ and scalar output $\boldsymbol{y}$.}
  \label{fig:hidden-act}
\end{figure}
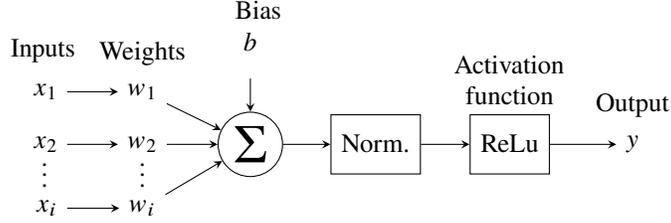

Hidden units contain a linear combination of the input vector with bias and are fed to a normalization layer implemented according to \cite{ba2016layer}. This is to reduce the difficulty of training networks whose inputs have different scales and non-zero means. The normalized value is then given to a ReLu activation function, as seen in Fig.~\ref{fig:hidden-act} for a generic hidden unit. The output layers, on the other hand, only contain a linear combination of the second hidden layer units with bias. All weights, biases and normalization factors are contained in network parameter vectors $\textbf{k}$ and $\boldsymbol{\theta}$ for the Q-function and the policy, respectively.

\vspace{-5mm}

\subsection{Experiment Setup}
\label{subsec:setup}

In this research, the \sac controllers are trained offline on the normal plant dynamics and their adaptation to failures is subsequently evaluated online based on their robust response. For comparison purposes, the adaptive response to failures is also generated.

\vspace{-1mm}
\subsubsection{Offline Learning}
\label{subsec:setup-ol}

\begin{figure}[hbt!]
    \centering
    \includegraphics[width=.6\textwidth]{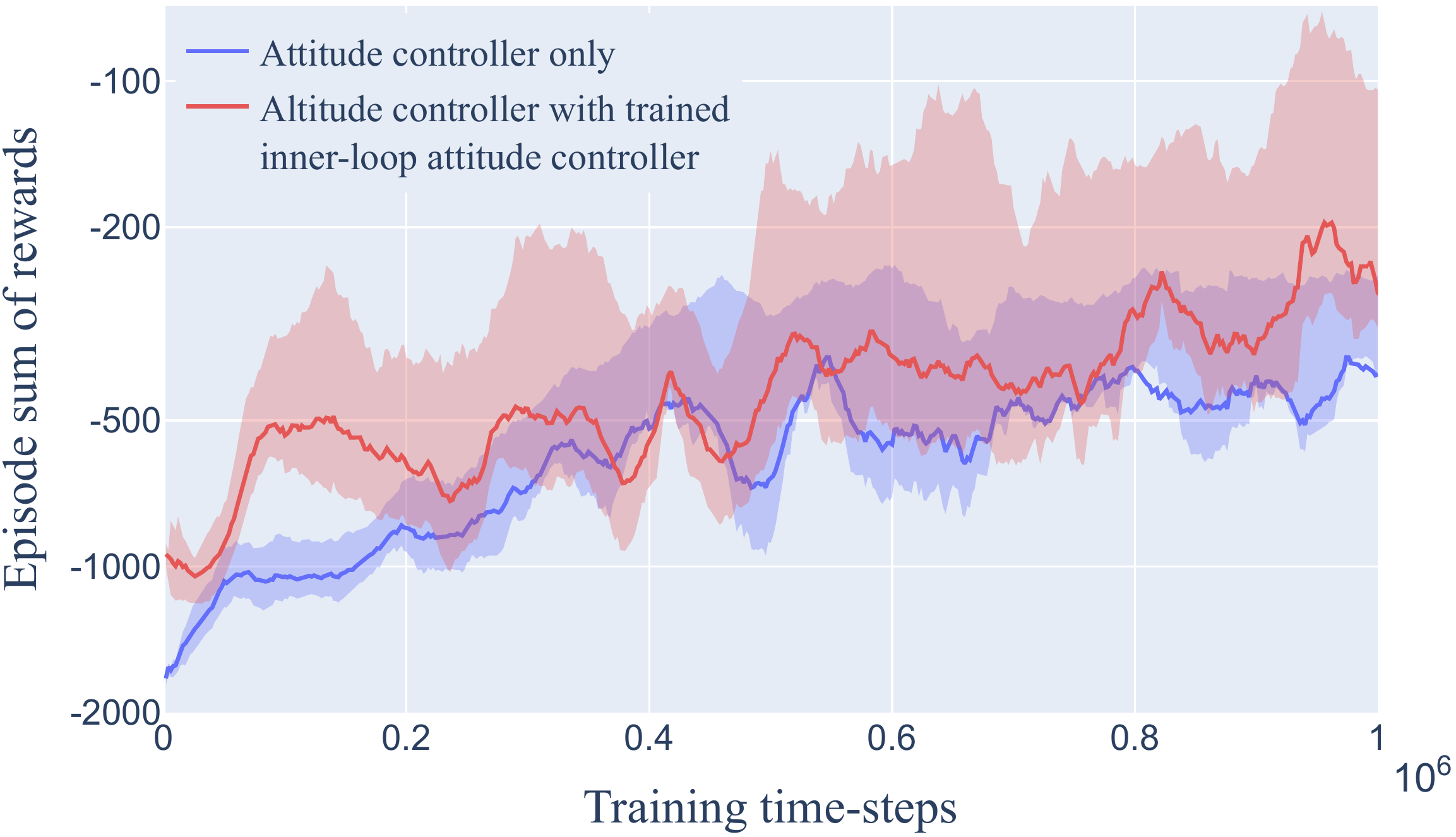}
    \caption{Sum of rewards in function of training time-steps. The curves show the mean (smoothed with a window of length 20) and the shaded region the interquartile range over 5 random seeds of successful trials.}
    \label{fig:training}
\end{figure}

\acrshort{DRL} agent training is best performed offline as it typically requires $10^6$ time steps or more. Normal plant dynamics with initial altitude and speed of $2000$m and $90\text{ms}^{-1}$, respectively, are used for the entire training process. For better learning stability, the inner-loop attitude controller is trained first alone with step reference signals for the pitch and roll angles, while the sideslip reference is always zero. After 500 $20$s-training episodes, or $10^6$ time steps, the positive learning curve in Fig.~\ref{fig:training} reaches a plateau, suggesting that no further training is required.

The altitude controller training is subsequently performed with the fully trained attitude controller in the inner loop. Successive climbing, constant altitude, and descending tasks allow the controller reach a converged policy after $10^6$ time steps. As observed in Fig.~\ref{fig:training}, training is quite unstable as significant performance drops are experienced even in the last training stages. Benefiting from the offline learning environment, training for both controllers is repeated until a satisfactory policy is reached.\\

\vspace{-3mm}

\vspace{-6mm}

\subsubsection{Online Robust Adaptation}

The adaptation of the cascaded \sac controller to unseen flight conditions, atmospheric disturbances and sensor noise is evaluated with its robust response, as no controller parameters are changed. Similarly, six unknown failure cases are simulated online and the controller has to robustly adapt its response without changing its parameters.

As an additional experiment, the \sac attitude controller is also trained on the failed system. While this experiment falls outside the scope of the research objective since simulation models of failed dynamics are typically not obtainable, it provides insights into possible performance improvements. The previously-described offline training process of the attitude controller is repeated for each failure case to obtain six adaptive agents. The adaptive response will, however, not be used to assess the fault-tolerance of the \sac controller.

\section{Results and Discussion}
\label{sec:ana}

The Cessna Citation jet aircraft response with the \sac controller is evaluated in this section, both on the non-failed and failed system. Moreover, the effect of biased sensor noise and atmospheric disturbances on the response is assessed. Lastly, additional robustness and reliability tests are conducted.

\subsection{Non-Failed System}

\begin{figure}[H]
    \centering
    \includegraphics[width=1\textwidth]{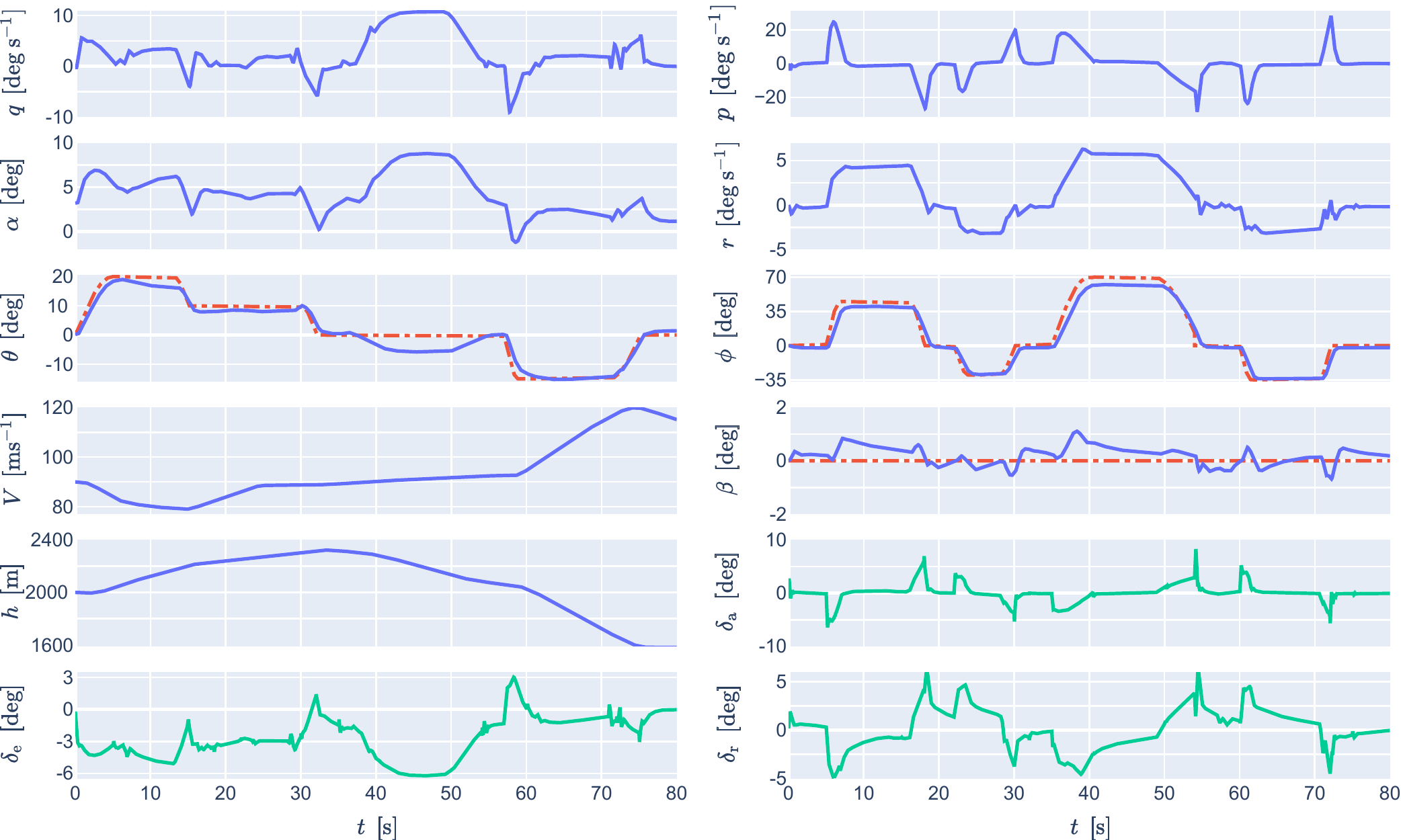}
    \caption{Attitude tracking response with \sac attitude controller. Reference signals are shown with red dashed lines and control inputs with green solid lines.}
    \label{fig:resp-nonfailed-att}
\end{figure}

One of the goals of this experiment is to show that the controller can operate the non-failed aircraft on a representative coupled attitude tracking task. As depicted in Fig.~\ref{fig:resp-nonfailed-att}, the attitude controller alone is able to keep the aircraft stable and minimizes the error overall. As the aircraft is initially untrimmed, rapid variations in the control input is seen close to $t=0$s. A large roll angle is reached despite remaining $7.5^\circ$-off from the $70^\circ$ reference, as the controller finds a compromise with the pitch angle error given the difficulty of keeping the aircraft horizontal during such a large bank maneuver. The roll angle suffers from a small yet consistently positive steady-state error, which could be due to the controller having failed to completely learn the asymmetricity of the aircraft.

The response of a similar aircraft on an analogous roll angle task in \cite{ferrari2004online} with a model-based \acrshort{DHP} controller shows less transient-response damping and a sideslip angle up to $50$ times larger. The velocity, on the other hand, is better tracked there since it is directly controlled by the \acrshort{DHP} agent, unlike here with an external auto-throttle.\\

\begin{figure}
    \centering
    \includegraphics[width=1\textwidth]{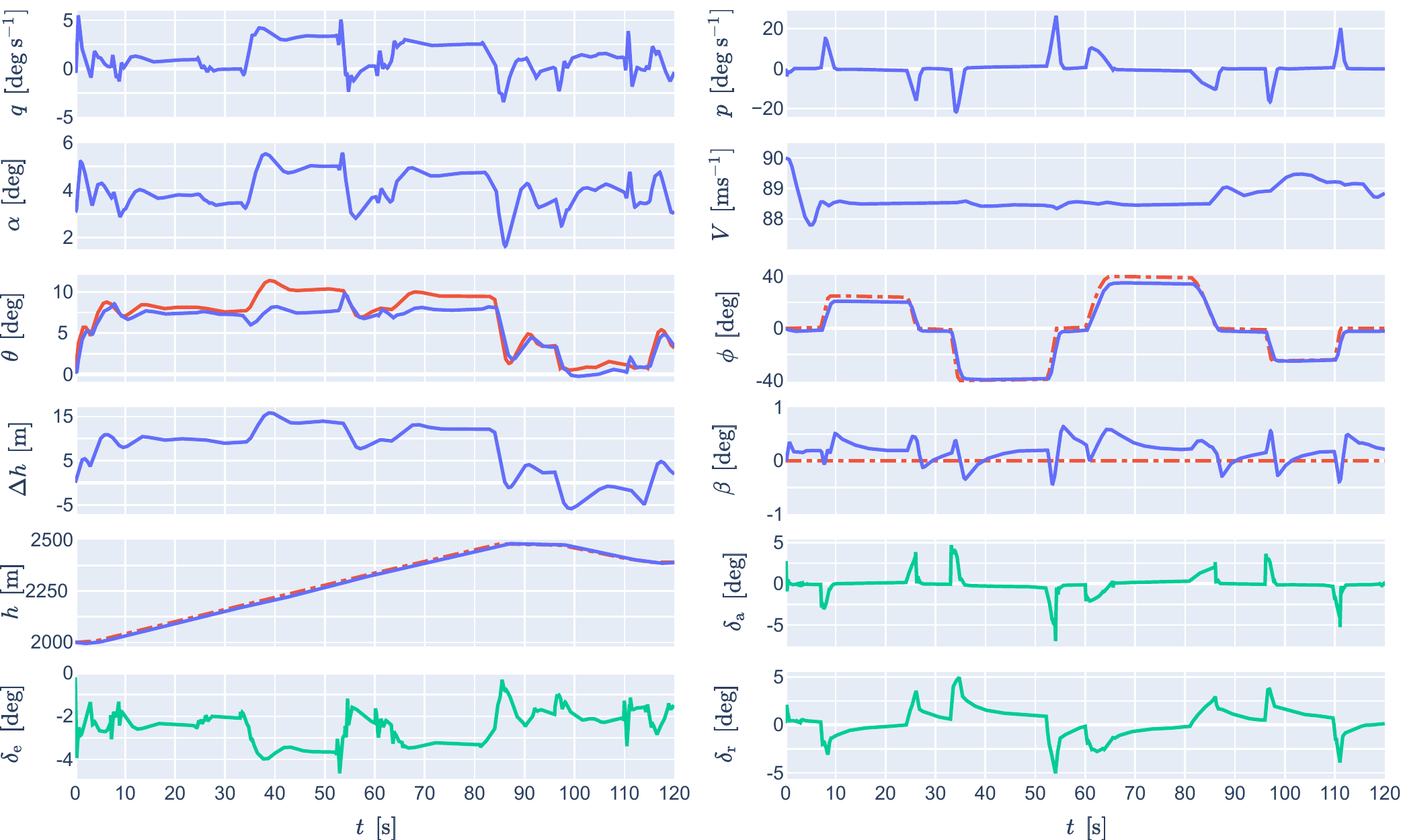}
    \caption{Altitude tracking response with cascaded \sac controller. External and self-generated reference signals are shown with red dashed and solid lines, respectively, and control inputs with green solid lines.}
    \label{fig:resp-nonfailed-alt}
\end{figure}

The proposed framework should also be able to complete a representative altitude tracking task with the cascaded controller structure. As seen in Fig.~\ref{fig:resp-nonfailed-alt}, successive climbing turns are tracked well, with an altitude error of $15$m or lower. This confirms that expert knowledge in flight control is successfully used to integrate the inner-loop and outer-loop controllers. Coupling effects in the attitude controller are observed as the elevator is pushed further up during the $40^\circ$ bank turns at $t=35$s and $t=65$s, which is not sufficient to track the altitude controller's higher pitch angle reference. This could be due to the attitude controller avoiding regions of pitch rate higher than $5^\circ$/s as long as the pitch error is not too large, unlike in Fig.~\ref{fig:resp-nonfailed-att} where it reached $6^\circ$. 

Similar performance to the Cessna Citation 500 response with uncoupled \acrshort{IDHP} agent in \cite{heyer2020online} is achieved, this time with a maximum roll angle of $40^\circ$ instead of $25^\circ$, benefiting from the coupled-dynamics \sac attitude controller.

\subsection{Failed System}

Another goal of this research is to evaluate the cascaded \sac controller on several types of failures. The robust response is shown first as it experiences an unexpected change in plant dynamics not seen during training. To identify possible performance gains, it is followed by the adaptive response, corresponding to a \sac controller trained on the failure case. For this research, however, the robust response is most important to determine if the \sac controller is successful since a plant model for every failure is typically not available for offline training.

\subsubsection{Jammed Rudder}

\begin{figure}[H]
    \centering
    \includegraphics[width=1\textwidth]{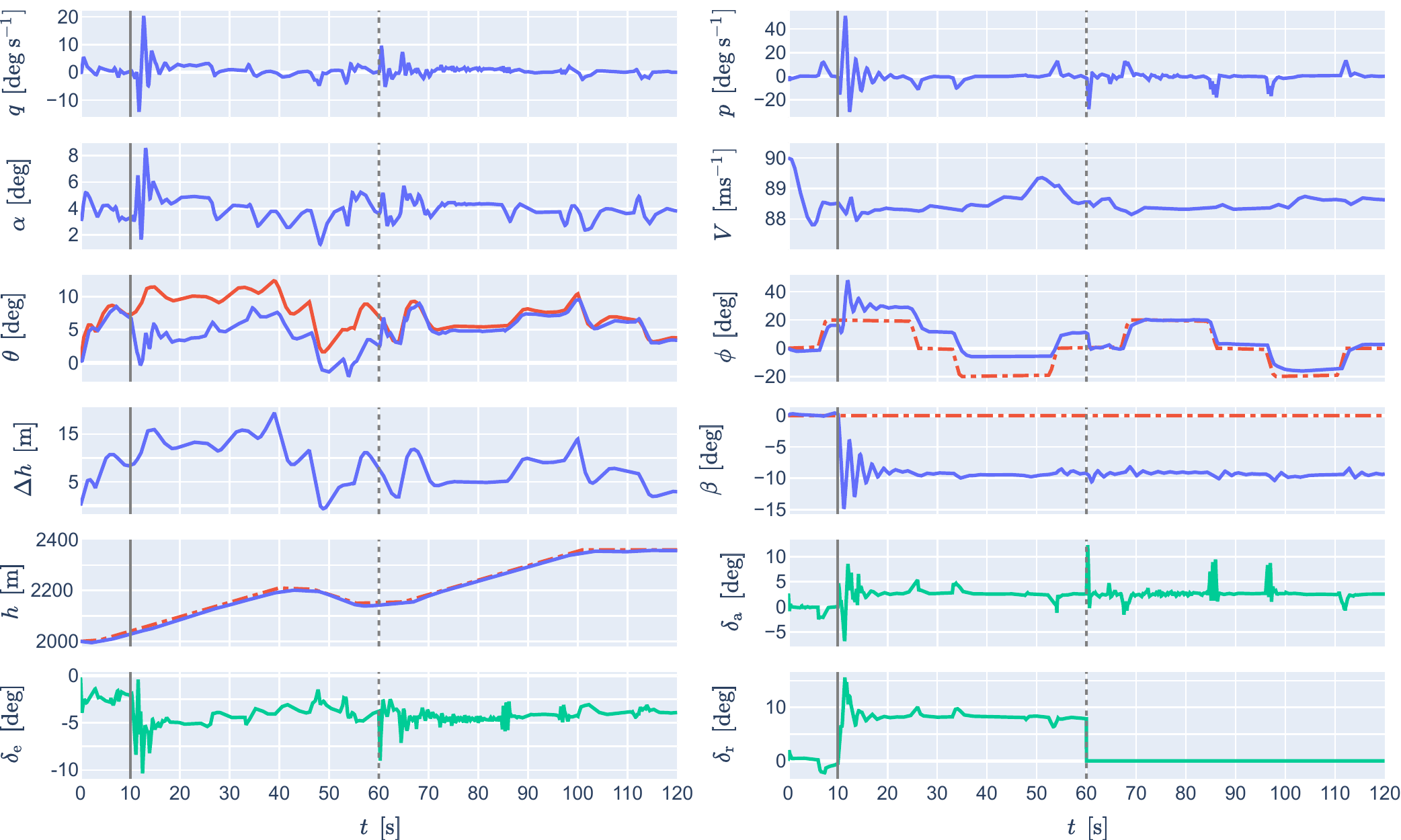}
    \caption{Altitude tracking response with rudder stuck at $\boldsymbol{\delta_\text{r}=-15^\circ}$ from $\boldsymbol{t=10}$s (grey solid line). External and self-generated reference signals are shown with red dashed and solid lines, respectively, and control inputs with green solid lines. Robust control until $\boldsymbol{t=60}$s, adaptive control thereafter (grey dotted line).}
    \label{fig:resp-dr}
\end{figure}

The response on a failure case with the rudder struck at $\delta_\text{r}=-15^\circ$ from $t=10$s onward is displayed in Fig.~\ref{fig:resp-dr}. The robust response until $t=60$s is stable despite the severe failure. The sideslip error is inevitably big with this rudder deflection, but large and relatively constant $15^\circ$ and $5^\circ$ errors in the roll and pitch angles, respectively, are also observed. The attitude controller is affected by the unusual dynamics, resulting in large body rates oscillations at the failure time. They are nevertheless contained as it starts operating the elevator and ailerons around a new offset position, and as expected, it also tries to deflect the rudder in the opposite direction to the one it is stuck in, with no effect. The altitude controller, on the other hand, manages to track the climb task successfully by producing a higher-than-normal pitch reference, thereby almost removing the effect of the large pitch angle error. Overall, the robust controller leverages its knowledge of coupling effects jointly using the elevator and ailerons around a new offset to counteract the rudder failure.

The adaptive response was obtained by training the controller on the failed system and by stopping it both from tracking the sideslip and controlling the jammed rudder. The response then exhibits a tracking performance similar to the non-failed case in the pitch and roll angles even though the sideslip angle error is still large.\\

A similar failure case on a business-jet aircraft with a model-dependent \acrshort{DHP} controller with the rudder stuck at $\delta_\text{r}=-15^\circ$ and an asymmetric loss of thrust (possibly beneficial for this failure case) was investigated in \cite{ferrari2004online}. The adaptive response was stable but showed large $10^\circ$-amplitude undamped sideslip angle oscillations and a roll angle error of up to $50^\circ$. Oscillations in the longitudinal plane had five times higher amplitude and two times higher frequency than the robust controller in Fig.~\ref{fig:resp-dr}. Overall, the \sac robust response is more stable with smaller oscillations and lower error than the \acrshort{DHP} adaptive one. This is partly attributable to the high generalization power of \acrshort{DNN}s and the robustness of stochastic policies.

\subsubsection{Reduced Aileron Effectiveness}

\begin{figure}
    \centering
    \includegraphics[width=1\textwidth,height=\fithere]{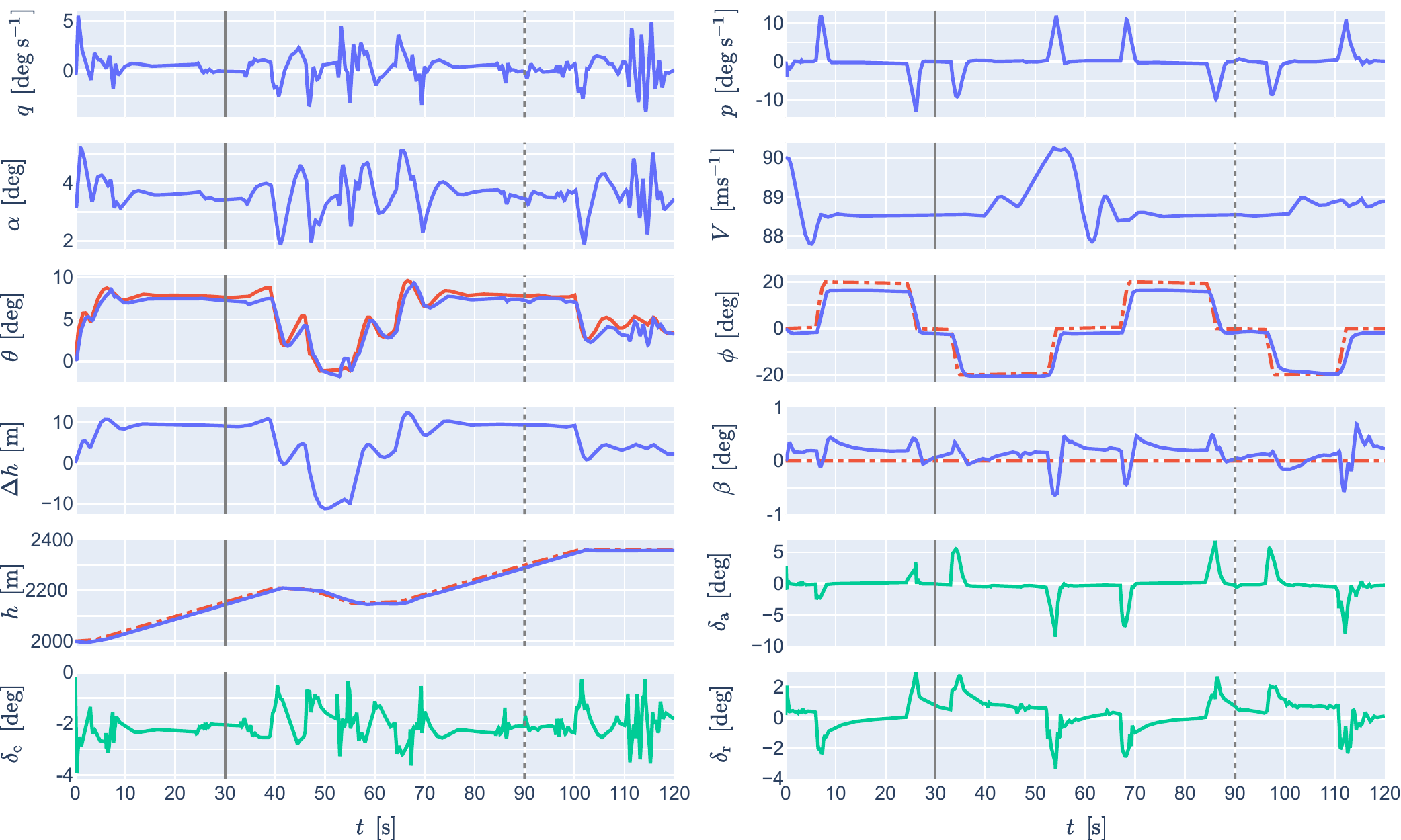}
    \caption{Altitude tracking response with 70$\boldsymbol{\%}$ reduced aileron effectiveness from $\boldsymbol{t=30}$s (solid grey line). External and self-generated reference signals are shown with red dashed and solid lines, respectively, and control inputs with green solid lines. Robust control until $\boldsymbol{t=90}$s, adaptive control thereafter (grey dotted line).}
    \label{fig:resp-da}
\end{figure}

An aileron failure case, implemented as a $70\mathbf{\%}$ reduced aileron effectiveness, is depicted in Fig.~\ref{fig:resp-da}. The aileron failure has few effects on the response. Most notably, the robust controller more than doubles and prolongs non-zero aileron deflections to track the roll angle reference with almost no difference compared to the non-failed plant. The roll rate decreases by up to $6\%$ with respect to the non-failed value.

The adaptive response from $t=90$s exhibits almost identical performance, although its control policy is reduced in magnitude resulting in a slightly slower transient response in the roll angle tracking. It also suffers from some oscillations in the longitudinal plane, making its response less desirable than the robust controller. The lower performance likely comes from the increased difficulty to learn on the failed system.\\

In \cite{heyer2020online}, an aileron failure with a $50\%$ reduced effectiveness was introduced on a Cessna Citation 500 with an \acrshort{IDHP} controller. Benefiting from its adaptive capability, the response was stable and well tracked in spite of the failure. It is worth noting that the robust \sac controller showed equivalent performance on a more severe failure, indicating again the high robustness of stochastic policies.

\begin{figure}
    \centering
    \includegraphics[width=1\textwidth,height=\fithere]{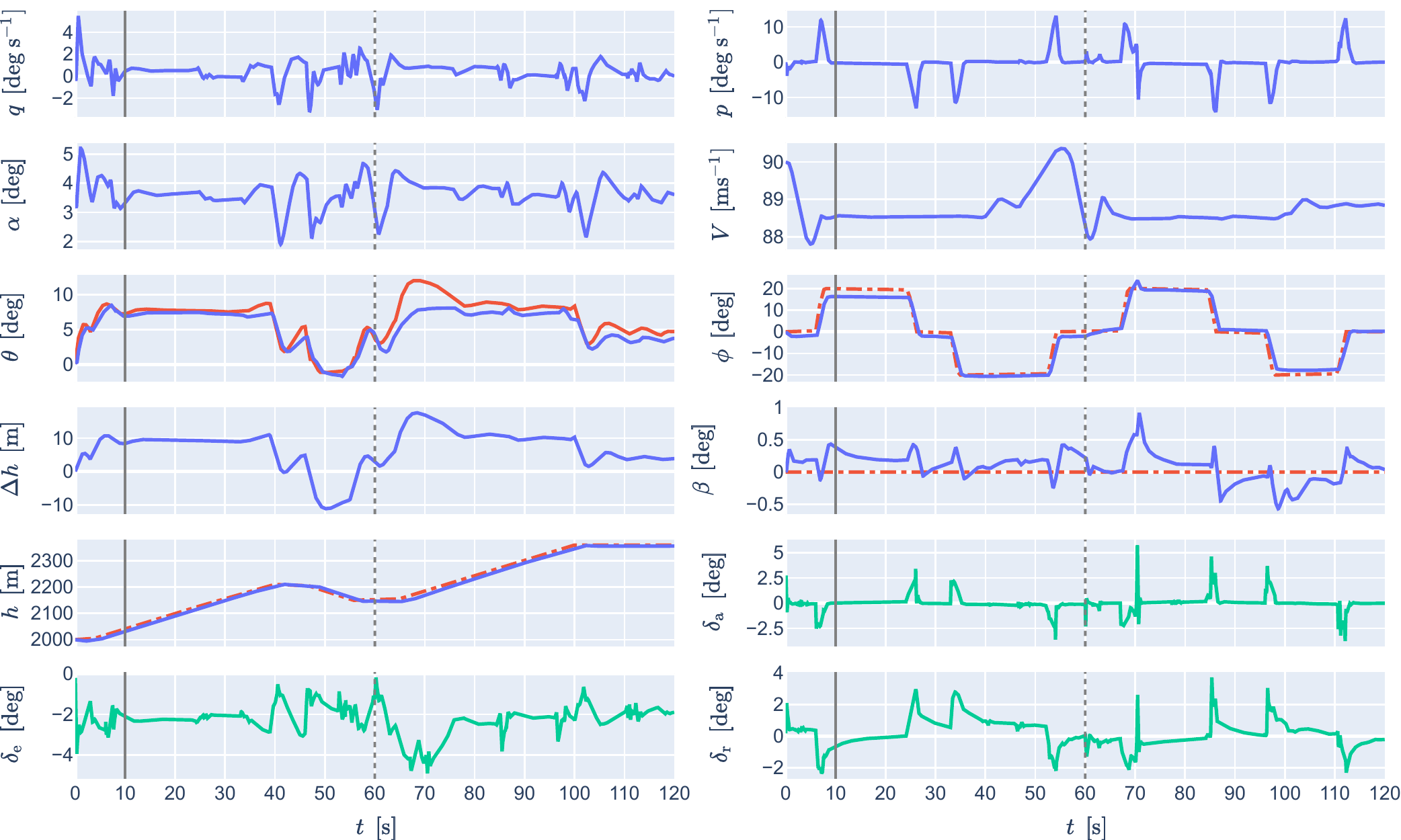}
    \caption{Altitude tracking response with reduced elevator range from $\boldsymbol{t=10}$s (grey solid line). External and self-generated reference signals are shown with red dashed and solid lines, respectively, and control inputs with green solid lines. Robust control until $\boldsymbol{t=60}$s, adaptive control thereafter (grey dotted line).}
    \label{fig:resp-de}
\end{figure}

\subsubsection{Reduced Elevator Range}

Another actuator failure is tested, this time with the elevator range reduced from $[-20.05^\circ,14.90^\circ]$ to $[-2.50^\circ,2.50^\circ]$. The \sac agent is unaware of the failure and control inputs outside the new range are saturated. The robust response shown in \autoref{fig:resp-de} is little-affected by the failure from $t=10$s. While all states remain stable and well-tracked, the maximum pitch rate is reduced and remains at about half of the non-failed case seen in Fig.~\ref{fig:resp-nonfailed-alt}. This is explained by the close relationship between elevator deflection and pitch rate.

The adaptive response, from $t=60$s, degrades more because of the failure. Despite having been trained on this failure, it can be explained by the more difficult task of initiating the climb with the failure already present. The controller instructs for more than $10$s a control input outside the reduced elevator deflection range, which is not enough to avoid a large pitch error. The altitude error, in turn, doubles with respect to the non-failed case but the aircraft eventually completes the climb. It is worth noting that the agent did not explicitly learn the new saturation limits as its control input exceeds the bounds.

\subsubsection{Partial Loss of Horizontal Tail}

Structural failures are difficult to anticipate as they can affect various components of the aircraft. For this study, the structural failure of an essential part to flight control, the horizontal tail, is studied. As it mainly affects the elevator control effectiveness and pitch damping, it is implemented in the simulation model with a $70\%$ reduction in $C_{(L/D/m)_{\delta_{e}}}$ and $C_{m_{q}}$.

The failure at $t=10$s generates a sudden loss in elevator effectiveness, which the robust agent immediately counteracts by quadrupling the elevator deflection. Despite that, large pitch angle and altitude errors are observed as the achieved pitch rate is too low for the climbing task. Lateral motion states are unaffected by this failure. This failure case shows the ability of the agent to adapt to this unexpected structural failure by offsetting its control input.

The adaptive response from $t=60$s shows a similar control strategy with an unusually high elevator deflection while keeping most states well-tracked.

\begin{figure}[H]
    \centering
    \includegraphics[width=1\textwidth]{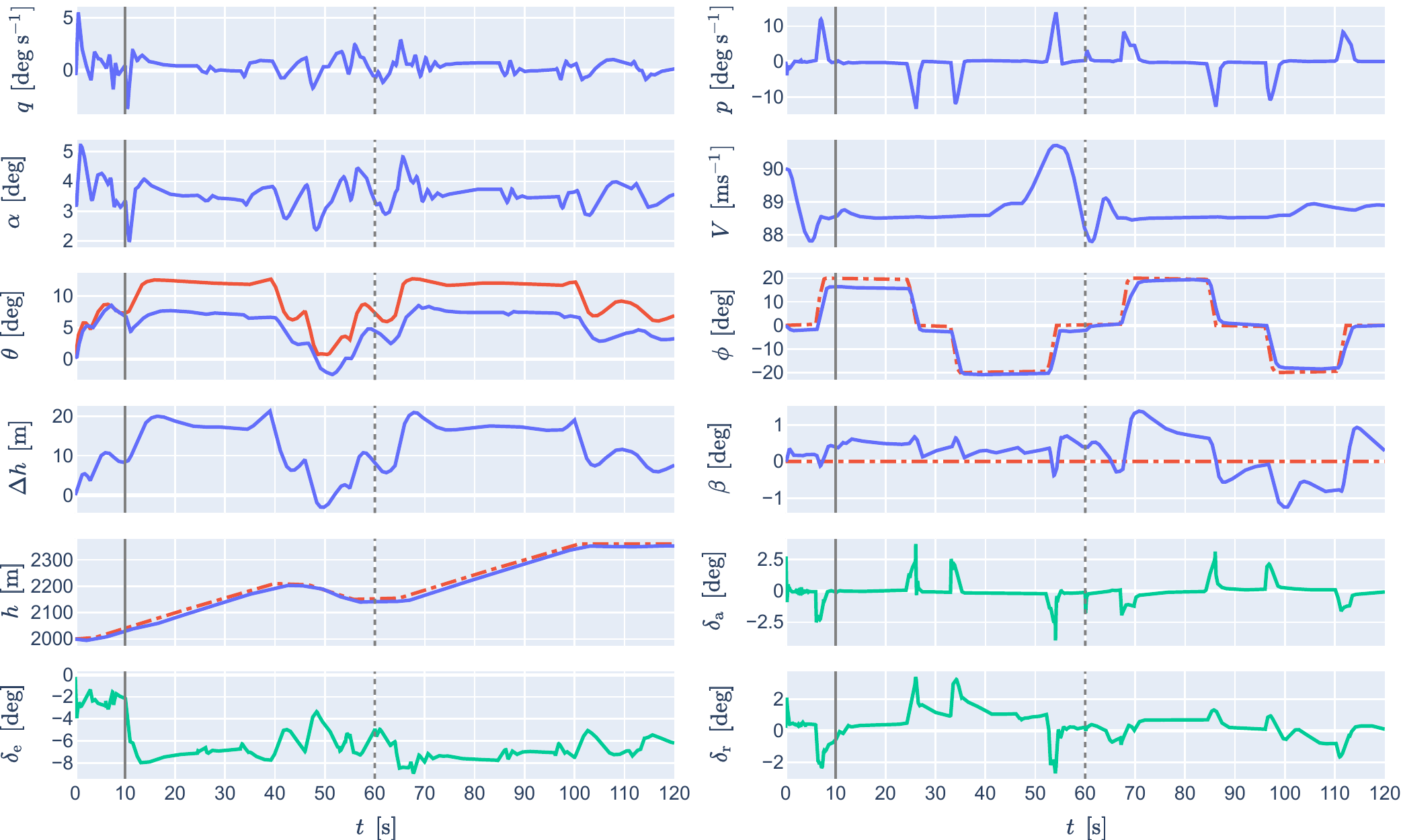}
    \caption{Altitude tracking response with partial loss of horizontal tail from $\boldsymbol{t=10}$s (grey solid line). External and self-generated reference signals are shown with red dashed and solid lines, respectively, and control inputs with green solid lines. Robust control until $\boldsymbol{t=60}$s, adaptive control thereafter (grey dotted line).}
    \label{fig:resp-ht}
\end{figure}

\subsubsection{Icing}

\begin{figure}[H]
    \centering
    \includegraphics[width=1\textwidth,]{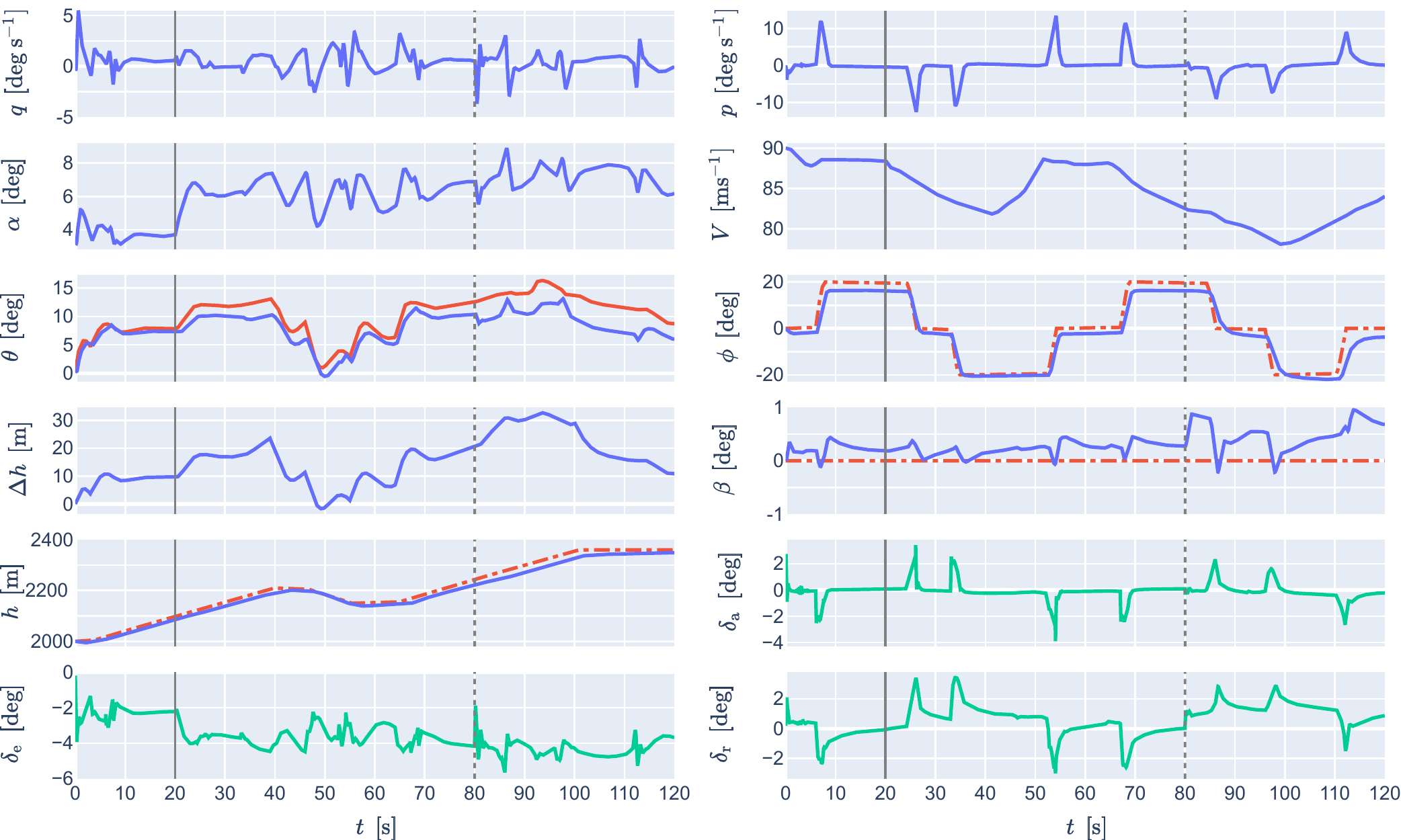}
    \caption{Altitude tracking response with icing from $\boldsymbol{t=20}$s (solid grey line). External and self-generated reference signals are shown with red dashed and solid lines, respectively, and control inputs with green solid lines. Robust control until $\boldsymbol{t=80}$s, adaptive control thereafter (grey dotted line).}
    \label{fig:resp-ice}
\end{figure}

Icing is a common challenge faced by aircraft as ice accretions accumulate on wings, which main effects include a reduction in maximum lift coefficient and an increase in drag coefficient \cite{lynch2001effects}. Conservative estimates suggest a reduction of $C_{L_{\text{max}}}$ by $30\%$ for mid-range Reynolds numbers and an increase of $C_D$ by $0.06$. As observed in Fig.~\ref{fig:resp-ice}, the decrease in lift and increase in drag due to icing cause large positive pitch and $20$m-altitude errors on the robust response. The robust controller still manages to achieve the climbing task by doubling its elevator deflection, although it avoids deflecting it further up than $-4^\circ$, most likely to avoid the stall region. The auto-throttle is unable to deal with both the increased drag and the climbing task, which leads to a reduction in velocity of $13\%$ compared to the non-iced plant. Lateral states, on the other hand, are unaltered by icing. 

The adaptive controller takes a more aggressive control strategy by deflecting the elevator further up, and despite trading kinetic energy, generates a larger altitude error. The roll angle transient response deteriorates and the sideslip error increases compared to the robust controller. It seems that the learning task with icing was difficult for the controller, which converged to a worse policy than the one of the robust controller.

\subsubsection{Center-of-Gravity Shift}

\begin{figure}[H]
    \centering
    \includegraphics[width=0.98\textwidth]{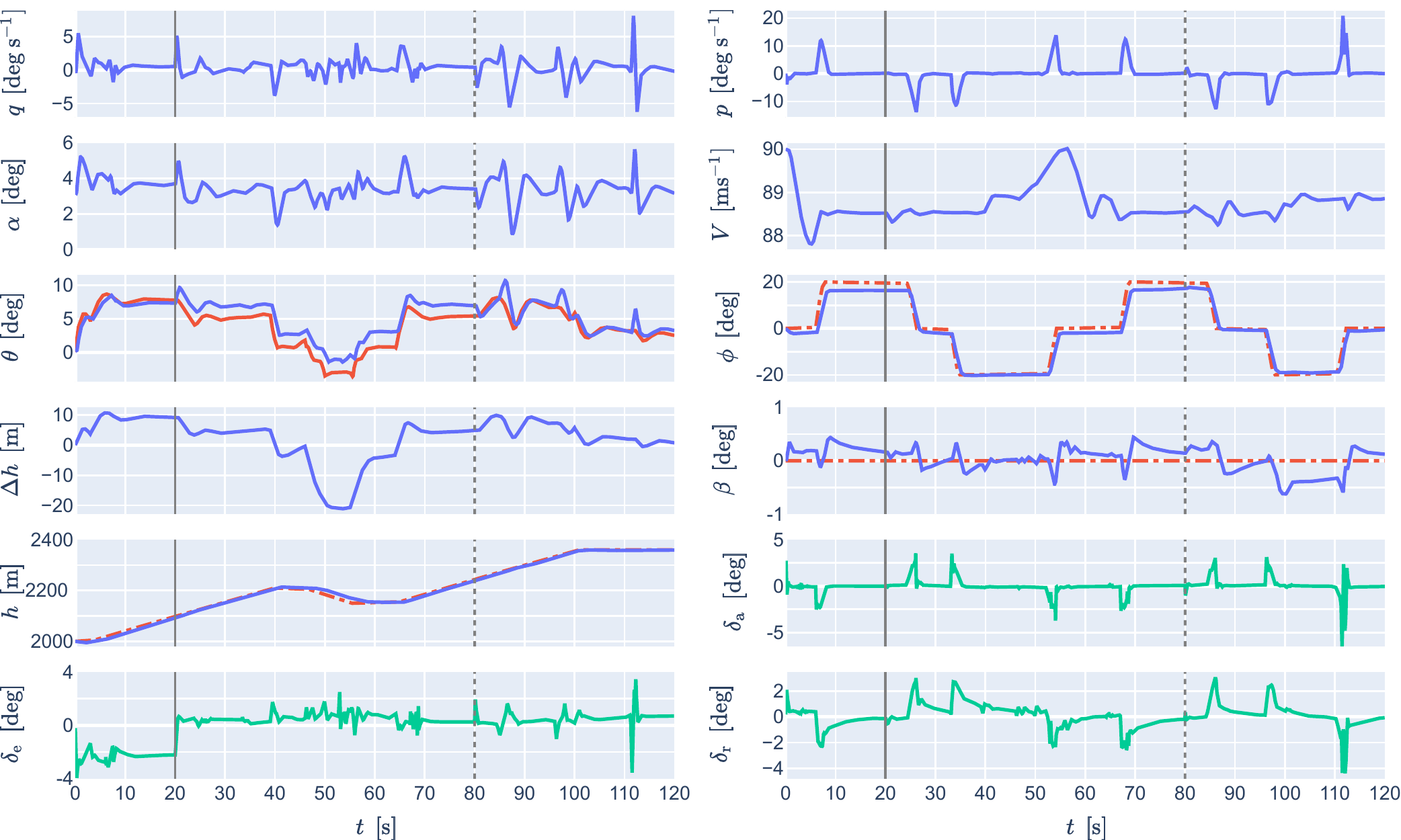}
    \caption{Altitude tracking response with c.g. shift at $\boldsymbol{t=20}$s (solid grey line). External and self-generated reference signals are shown with red dashed and solid lines, respectively, and control inputs with green solid lines. Robust control until $\boldsymbol{t=80}$s, adaptive control thereafter (grey dotted line).}
    \label{fig:resp-cg}
\end{figure}

A sudden backward shift of heavy cargo in the aircraft can cause a diminution of the stability margin and create instability. The event of a $300$kg payload moving from the front to the back of the passenger cabin is investigated, which translates to a backwards c.g. shift of $0.25$m on the PH-LAB. As seen in Fig.~\ref{fig:resp-cg}, 
the c.g. shift at $t=20$s causes a sudden change in the pitch rate and the controller immediately adapts the elevator deflection to the rare positive range. A constant negative pitch angle error appears as the c.g. moves backward, indicating that the controller's policy is stable but has not fully adapted to the new c.g. location. This is because the controller has only knowledge of the pitch error and not of the pitch angle itself, and has likely attributed higher elevator deflections to steep dive maneuvers on the aircraft with normal c.g. position. The robust attitude controller is unable to offset its elevator control input accurately. Consequently, the aircraft now has difficulty pitching down and experiences a large negative altitude error during descent. The lateral states, on the other hand, remain undisturbed by the c.g. shift.

The adaptive controller eliminates most of the pitch angle error, although it still suffers from sudden pitch-up events as the c.g. moved further back. This hints at instability issues should the c.g. be moved further back.

\subsection{Effect of Biased Sensor Noise and Atmospheric Disturbances}
\label{subsec:noise}

To evaluate the \sac controller in conditions closer to real-world phenomena, the addition of biased sensor noise and atmospheric disturbances is studied. A Gaussian white noise with standard deviation and bias obtained from the Cessna Citation PH-LAB's sensors in \cite{grondman2018design} are used to simulate sensor noise, with values shown in Table.~\ref{tab:noise}. Noise from control surface deflection measurements is disregarded as the attitude controller observes the control input instead. Atmospheric disturbances in the form of discrete vertical gusts ($15\text{ft}\:\text{s}^{-1}$), specified in MIL-F-8785C \cite{moorhouse1982background}, are also added to the system. They are implemented as $3$s-step disturbances on the angle-of-attack at $t=20$s and $t=75$s.

\begin{table}[H]
\centering
\caption{Cessna Citation PH-LAB aircraft sensor characteristics. Values from \cite{grondman2018design}.}
\label{tab:noise}
\begin{tabular}{lllll} 
\toprule
\textbf{Observed state} & $\boldsymbol{p,q,r \:[\textbf{rad}\:\textbf{s}^{-1}]}$ & $\boldsymbol{\theta,\phi\: [\textbf{rad}]}$ & $\boldsymbol{\beta\: [\textbf{rad}]}$ & $\boldsymbol{h \:[\textbf{m}]}$  \\
\cmidrule(r){1-5}
\textbf{Noise SSD}      &  $6.3\cdot10^{-4}$                      &      $3.2\cdot10^{-5}$                       &        $2.7\cdot10^{-4}$              &  $6.7\cdot10^{-2}$                \\
\textbf{Bias} & $3.0\cdot10^{-5}$ & $4.0\cdot10^{-3}$ & $1.8\cdot10^{-3}$ & $8.0\cdot10^{-3}$\\
\bottomrule
\end{tabular}
\end{table}

\begin{figure}[ht!]
    \centering
    \includegraphics[width=1\textwidth]{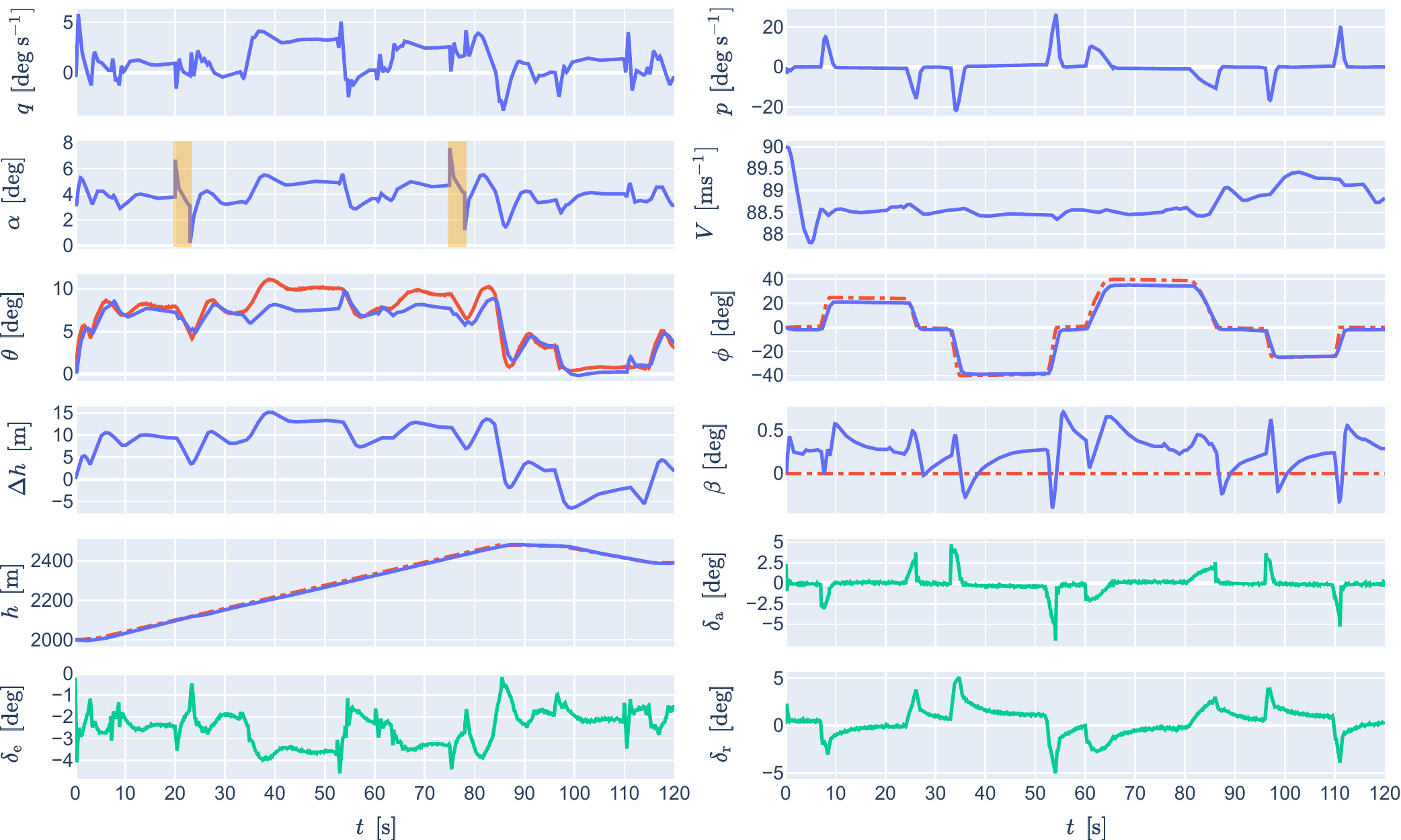}
    \caption{Altitude tracking response with biased sensor noise and $\boldsymbol{15\text{ft}\:\text{s}^{-1}}$ discrete vertical gusts (orange shaded region on $\boldsymbol{\alpha}$ plot). External and self-generated reference signals are shown with red dashed and solid lines, respectively, and control inputs with green solid lines.}
    \label{fig:resp-noise}
\end{figure}

As shown in Fig.~\ref{fig:resp-noise}, despite the controllers only having been trained with the ideal sensor assumption, the response remains stable and the errors are similar to the ideal-sensor scenario. The addition of biased sensor noise creates a slightly noisy control input and reference pitch angle while the aircraft states are unaffected.

The upwards vertical gusts cause a sudden increase in the angle-of-attack and disturb the pitch rate. As this makes the altitude error magnitude decrease, the outer-loop controller can instruct a lower pitch attitude, which subsequently leads to a downward elevator deflection. Conversely, when the gust vanishes, the sudden drop in angle-of-attack leads to a sharp upward elevator deflection to limit the altitude error. During this time, the lateral states are unperturbed. This response demonstrates the ability to reject atmospheric disturbances in the presence of biased sensor noise.

\subsection{Additional Tests}

Further tests in terms of robustness and training reliability are conducted in this section, in an effort to explore some limitations of the \sac controller.

\subsubsection{Robustness Analysis}

The cascaded SAC controller was evaluated so far on the same nominal \acrfull{IFC} and similar reference signal shapes. It is interesting to investigate how general the controller is by evaluating its robustness to \acrshort{IFC} it has not been trained on. Initial altitude and speed combinations are chosen to match previous tests done in \cite{heyer2020online,leethesis} on the same aircraft. The robustness to various reference signal shapes and frequencies is also evaluated, namely on sinusoidal and triangular reference signals for the altitude and roll angle reference signals, while the sideslip reference angle remains zero. A performance metric is established as the average \acrfull{nMAE} over externally tracked states (altitude, roll and sideslip angles). Normalization is done over the reference signals range, while for the zero-referenced sideslip angle an acceptable range of $[-5^\circ,5^\circ]$ is used. 

As reported in Table~\ref{tab:relia}, the controller is robust on all four \acrshort{IFC}, where the largest error corresponds to the condition with the lowest dynamic pressure. A higher flight speed leads to an increased performance while a higher altitude leads to a decreased one, explained by higher and lower dynamic pressure and aerodynamic damping, respectively. Different reference signal shapes are tracked with a \acrfull{nMAE} no higher than $5\%$, with the error mostly caused by the slower transient response of altitude on the high-frequency signals. Overall, the controller is observed to maintain a relatively low error on various \acrshort{IFC} and a multitude of reference signal shapes.

\begin{table}
\centering
\caption{Robustness analysis to varying \acrshort{IFC} and reference signal shapes. The controller was trained on the nominal \acrshort{IFC}.}
\label{tab:relia}
\begin{tabular}{llll} 
\toprule \textbf{Reference signals $\boldsymbol{h^R,\phi^R}$} & \textbf{Initial altitude} $\boldsymbol{[\textbf{m}]}$ & \textbf{Initial speed} $\boldsymbol{[\textbf{m}\textbf{s}^{-1}]}$ & \textbf{\acrshort{nMAE}}   \\
\cmidrule(r){1-4}
As in Fig.~\ref{fig:resp-nonfailed-alt}    &  $2000$ (nominal) & $90$ (nominal)        &  $2.64\%$                                          \\
As in Fig.~\ref{fig:resp-nonfailed-alt}    &  $2000$ & $140$        &      $1.95\%$                                   \\
As in Fig.~\ref{fig:resp-nonfailed-alt}$^*$    &  $5000$ & $90$         &     $3.16\%$                                  \\
As in Fig.~\ref{fig:resp-nonfailed-alt}$^*$    &  $5000$ & $140$        &    $2.13\%$                                   \\
Sinusoidal (low frequency)   &  $2000$ &$90$   &            $3.22\%$                                           \\
Sinusoidal (high frequency)   &  $2000$ &$90$   &            $5.00\%$                                           \\
 Triangular (low frequency)                      &     $2000$ &$90$&   $3.00\%$                                                               \\
  Triangular (high frequency)                      &     $2000$ &$90$&   $4.76\%$                                                               \\
\bottomrule
\multicolumn{4}{l}{\footnotesize low frequency: $T_{h^R}=80\text{s},A_{h^R}=80\text{m},T_{\phi^R}=50\text{s},A_{\phi^R}=50^\circ$} \\
\multicolumn{4}{l}{\footnotesize high frequency: $T_{h^R}=40\text{s},A_{h^R}=40\text{m},T_{\phi^R}=25\text{s},A_{\phi^R}=25^\circ$} \\
\multicolumn{4}{l}{\footnotesize $^*h^R$ is offset to new initial altitude} \\
\end{tabular}
\end{table}

\subsubsection{Training Reliability Analysis}

Training the \sac controllers involves several random processes, including sampling from the stochastic policy to choose actions, parameter initialization for the \acrshort{DNN}s and minibatch sampling. Since the performance can vary from training to training, it is interesting to evaluate how reliable the training process is. Although a large enough number of trials can be granted in an offline learning environment, a high success rate would indicate the convenience of such process. 

Samples are generated by training pairs of cascaded SAC controllers in the same conditions as described in \autoref{subsec:setup-ol}. Each sample is evaluated on the nominal \acrshort{IFC} and altitude tracking task shown in Fig.~\ref{fig:resp-nonfailed-alt} with a $5\%$ \acrshort{nMAE} success threshold. A global training success rate is obtained by generating enough samples ($n=27$) until convergence to a stable success rate is obtained. It is found to be at $26\%$. This low value reveals the difficulty to train with consistency stochastic policy-based and random initialization-dependent \acrshort{DRL} algorithms. The high sample efficiency, generalization power and robustness to failures presented so far can be seen to come at the expense of learning stability. It should be noted, however, that this does not compromise the online reliability and performance of the SAC controller, as long as it is trained in an offline environment.

\section{Conclusion}
\label{sec:ccl}

A controller employing \acrfull{SAC} \acrfull{DRL} for the coupled-dynamics control of the Cessna Citation 500 jet aircraft is built as part of this research. Expert knowledge from classic flight control is used to create a cascaded \sac controller structure. The response is stable and keeps the tracking error low to successfully achieve coordinated $40^\circ$-bank climbing turns and $70^\circ$-bank flat turns. Severe failures such as the rudder jammed at $-15^\circ$, the aileron effectiveness reduced by $70\%$ or other unforeseen events such as icing and c.g. shift are handled by the robust \sac controller such that stability is maintained and the tracking task is achieved, demonstrating the \sac controller's strong fault tolerance. This high performance is further exhibited on varying initial flight conditions and tracking task types, and on biased sensor noise and atmospheric disturbances, none of which considerably degrade the controller response. This is achieved not having experienced any of these unexpected changes in dynamics, tracking task or flight conditions during training, indicating a high robustness not achievable with model-based controllers. It is believed that \acrshort{SAC}'s stochastic policy and deep neural networks allowing for better exploration and a higher generalization power, respectively, are the main contributors to this ability.

It is seen, however, to come at the expense of a stable and consistent offline training performance, which makes the development of \sac controllers more difficult but does not jeopardize online performance. This low training reliability is mainly attributed to the various random processes of the \sac algorithm, from network initialization to its stochastic policy. Interestingly, when training on the failed system, i.e. switching from robust to adaptive control, the performance of the \sac controller worsens or remains the same on five out of six failure cases. It is explained by the increased difficulty of training on the failed plant dynamics, given the already low training reliability.

This research contributes to creating a coupled-dynamics model-free flight controller that allows for fault tolerance to various types of unforeseen failures. It is demonstrated that robust control through \acrshort{DRL} can, unlike model-based controllers, adapt to various normal and failed flight conditions. While \acrshort{DRL} is already widely used for small-scale UAV flight controllers, this research also paves the way for more applications to civil aircraft inner and outer-loop flight controllers. It is expected that \acrshort{DRL} methods will be used more broadly in flight control applications to increase fault tolerance and help achieve safe fully-autonomous flight. It is suggested for further research to explore deterministic policy-based \acrlong{TD3} or on-policy \acrlong{PPO} \acrshort{DRL} algorithms that, at the expense of enhanced exploration, can increase training reliability and may enable safe online learning, thereby removing the need for a plant model during offline learning.




\bibliography{sample}

\end{document}